\documentclass[usenatbib]{mnras}
\usepackage{newtxtext,newtxmath}

\usepackage[T1]{fontenc}

\DeclareRobustCommand{\VAN}[3]{#2}
\let\VANthebibliography\thebibliography
\def\thebibliography{\DeclareRobustCommand{\VAN}[3]{##3}\VANthebibliography}

\usepackage{graphicx}	
\usepackage{amsmath}	
\usepackage{booktabs}
\usepackage{hyperref}
\usepackage{makecell}
\usepackage{tabularx}
\usepackage{xcolor}
\definecolor{winered}{RGB}{114, 0, 38}
\usepackage{float} 
\usepackage{booktabs}
\usepackage{comment}
\usepackage{url}

\newcommand{\vel}{\mathrm{v}}

\def\be{\begin{equation}}
\def\ee{\end{equation}}
\def\bi{\begin{itemize}}
\def\i{\item}
\def\ei{\end{itemize}}
\def\ben{\begin{enumerate}}
\def\een{\end{enumerate}}

\title[Nucleosynthesis in fast ejecta]{Nucleosynthesis in the fast ejecta of a neutron star merger}

\author[L. Schnabel et al.]{
Lukas Schnabel,$^{1}$\thanks{E-mail: publications@ras.ac.uk (KTS)}
Stephan Rosswog,$^{1,2}$
Moritz Reichert$^{3}$
and Friedrich-Karl Thielemann$^{4,5}$
\\
$^{1}$Hamburg Observatory, Department of Physics, University of Hamburg, Gojenbergsweg 112, 21029 Hamburg, Germany\\
$^{2}$The Oskar Klein Centre, Department of Astronomy, AlbaNova, Stockholm University, SE-10691 Stockholm, Sweden\\
$^{3}$Departament d’Astronomia i Astrofísica, Universitat de València, C/Dr Moliner, 50, E-46100 Burjassot (València), Spain\\
$^{4}$Physics Department, University of Basel, Klingelbergstrasse 82, 4056 Basel, Switzerland\\
$^{5}$Theory, GSI Helmholtz Center for Heavy Ion Research, Planckstrasse 1, 64291 Darmstadt, Germany}

\date{Accepted XXX. Received YYY; in original form ZZZ}

\pubyear{\the\year{}}

\begin{document}
\label{firstpage}
\pagerange{\pageref{firstpage}--\pageref{lastpage}}
\maketitle


\begin{abstract}
Neutron star mergers are today considered a major production site for rapid neutron capture elements.  While the bulk of the matter escapes at fast, but non-relativistic velocities (${\sim} 0.2\,c$), a small amount of the dynamically ejected mass reaches mildly relativistic velocities (${\gtrsim}0.6\,c$). It has been suggested earlier, that in such ejecta parts neutrons may avoid being captured and that their decay could power an early blue precursor to the main kilonova event. Here we study in detail the nucleosynthesis in such fast ejecta with nuclear network calculations along both parametrized and numerical relativity trajectories. We find that the nucleosynthesis can be divided into three channels, in one of which a substantial amount of free neutrons survives when  the main r-process has frozen out. We provide a (semi-)analytical model for surviving free neutrons which agrees very well with the network calculations. If the mass fraction of the free neutrons exceeds ${\sim} 0.05$, their $\beta^-$-decay dominates the nuclear heating rate between ${\sim} 100$ and ${\sim} 10^4$ seconds. This dominance leads to a pronounced kilonova precursor that should for plausible ejecta parameters be visible for \textit{ULTRASAT} out to ${\sim}200\,\rm Mpc$. Since at low electron fractions free neutrons can survive even for moderate velocities, mergers with large tidal ejecta, such as asymmetric neutron star mergers or favorable neutron star black hole mergers,
may produce particularly bright blue precursors to their subsequent kilonovae.
\end{abstract}




\section{Introduction}
\begin{figure*}
  \centering
  \includegraphics[width=0.8\textwidth]{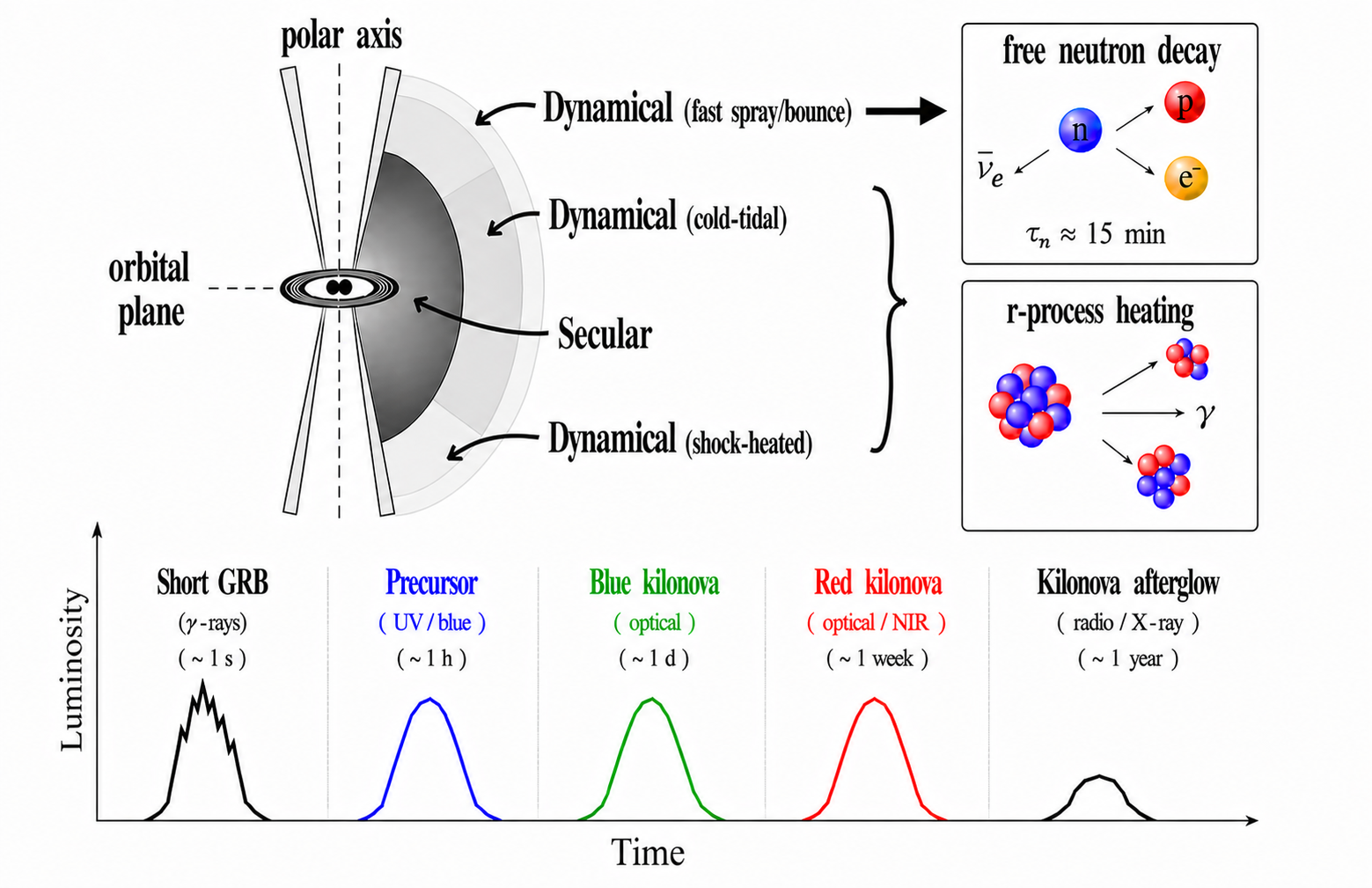}
  \caption{Schematic overview of the ejecta structure and resulting electromagnetic emission from a neutron star merger. The top panel shows the angular distribution of dynamical and secular ejecta (sketch not to scale), together with their dominating heating source. The bottom panel shows the electromagnetic counterparts, along with their typical timescales and the wavelength bands where they can be observed.}
  \label{fig:ejecta_overview}
\end{figure*}

The dynamics of a neutron star merger (NSM) is governed by physics at the extremes \citep{shibata16,baiotti17,sarin20}.
The neutron stars strongly and dynamically curve their surrounding spacetime \citep{alcubierre08,baumgarte10,rezzolla13a}, 
their  densities substantially exceed nuclear matter density 
and in the post-merger phase the temperatures surpass the core temperature of the Sun by more than four orders of magnitude \citep[e.g.][]{perego19a}. Under these conditions, the neutron star matter is no longer 
in $\beta$-equilibrium and the weak interactions become so fast that neutrinos
are emitted at luminosities exceeding $10^{53}$ erg/s \citep[e.g.][]{rosswog03a,sekiguchi11,foucart23}. Moreover, initial "seed" magnetic
fields are massively amplified to strengths that can locally exceed $\sim10^{17}$ G \citep[e.g.][]{price06,ruiz16,kiuchi15,kiuchi18,kiuchi24,aguilera24,Aguilera25,cook26,neuweiler26}. Clearly,
terrestrial experiments cannot even get close to these extreme conditions, the only way to probe the involved physics is therefore the comparison of theoretical 
predictions -- ideally for several  messengers -- with potential observations.\\
Among the known signatures of neutron star mergers are gravitational waves, short (and potentially even long) 
gamma-ray bursts (GRBs) \citep[e.g.][]{piran04,meszaros06,nakar07,lee07,kumar15,gottlieb23} and "kilonovae" \citep[e.g.][]{li98,kulkarni05,rosswog05a,metzger10b,roberts11,rosswog18a,wu19,metzger20,kawaguchi24} powered by the radioactive decay of freshly 
synthesized heavy elements. While kilonovae are the result of the bulk of the neutron
star ejecta that undergo ''rapid neutron capture'' or r-process \citep{Freiburghaus1999,thielemann26}, there is a small fraction
of the ejecta that is so fast that the initially free neutrons can decay before they are
captured by nuclei. This scenario was initially discussed by \cite{kulkarni05}, and specifically for fastest part of NSM ejecta, by \cite{metzger15a}.\\
With mean lifetimes of only a few minutes, the decaying neutrons in these fast outflows could cause an early blue precursor to the main kilonova event 
\citep{metzger15a}, and -- after many months -- also cause a ''kilonova afterglow'' due to synchrotron emission \citep{nakar11a,mooley17,hotokezaka18a,hajela22,sadeh23,Sadeh24}. While small amounts of fast ejecta have been seen in a number of simulations with different methodology \citep[e.g.][]{hotokezaka13,bauswein13a,kyutoku14,kiuchi17,radice18a,hotokezaka18a,dean21,nedora21,rosswog22b,combi23}, their origin was not entirely clear.
A recent study \citep{rosswog25b} with the Lagrangian Numerical relativity code \texttt{SPHINCS\_BSSN} \citep{rosswog21a,diener22a,rosswog23a,rosswog25c} found that such fast ejecta are launched in several pulses. The first pulse is the result of matter being ''sprayed out'' from the interface between
the neutron stars. This matter expands predominantly in the orbital plane.
Subsequently, several pulses are launched 
when the central merger remnant is deeply compressed and then ''bounces back''. These ''bounce'' pulses are close to spherically symmetric and the first such pulse is often the fastest and can therefore catch up with the first-launched "spray pulse".\\
In addition to these shock-driven fast ejecta, there are also indications that binaries with large mass ratios, either two neutron stars or a neutron star black hole system, can accelerate ejecta to large velocities via tidal torques, see e.g.\ \cite{matur24} and \cite{bernuzzi25}. \\
While the launch mechanism
has become clearer, the exact amounts of fast ejecta and their detailed properties are still extremely hard to resolve. It is, however, clear that with higher numerical resolution one finds faster ejecta, potentially  reaching,
for small amounts, close to the speed of light. Moreover, softer equations of state (EOS) lead to larger peak velocities within the ejecta, indicating that fast ejecta signatures may carry the imprints of the incompletely understood high-density equation of state. A schematic summary of the ejecta components and their electromagnetic signatures is shown in Fig.~\ref{fig:ejecta_overview}.\\
While it is qualitatively plausible that for fast enough neutrons the density drops
so rapidly that they can avoid being captured by nuclei, we want to study here the nucleosynthesis within fast ejecta in more detail. In Sec.~\ref{sec:network} we concisely summarize the ingredients and methods of the nuclear reaction network that we are using in this study, and in Sec.~\ref{sec:trajectories} we summarize how we construct our parametrized trajectories. We also compare against trajectories that we find in our numerical relativity simulations which we briefly summarize in Sec.~\ref{sec:NSNS_simulations}. In Sec.~\ref{sec:kn_model} we describe how we calculate light curves and in Sec.~\ref{sec:results} we present the main results of our study before  we summarize our findings in Sec.~\ref{sec:summary}.

\section{The nuclear reaction network}
\label{sec:network}
For our nucleosynthesis calculations, we use the open-source nuclear reaction network \texttt{WinNet} \citep{winteler12, reichert23}, including approximately 6\,700 nuclei. At temperatures above $9\, \mathrm{GK}$, we assume nuclear statistical equilibrium (NSE), using a combination of experimental and FRDM nuclear masses \citep{moeller95} from the JINA Reaclib database \citep{Cyburt2010}. We motivate our choice of the NSE temperatures in Appendix~\ref{sec:appendix_B}. Below $8.5\,\rm{GK}$, we solve the full nuclear reaction network equations, incorporating around 70\,000 reactions from the JINA Reaclib. In between $8.5$ and $9\, \rm{GK}$, we assume a transition phase in which the network can either be in NSE or in network mode. In both temperature regimes, we apply screening corrections following \cite{Kravchuk2014}. Since the weak reaction rates in the JINA Reaclib are not measured under stellar conditions, they are replaced above temperatures of $0.01\, \rm{GK}$ with theoretically calculated weak rates \citep{fuller95, Oda1994, langanke01, Pruet2003, Suzuki2016}. These include electron and positron captures, as well as $\beta$-decays on nucleons and nuclei. As the JINA Reaclib only provides experimental $\alpha$-decay rates, we supplement them with theoretical rates from \citet{Dong2005} and \citet{reichert23}. We also include spontaneous fission \citep{Khuyagbaatar2020}, neutron-induced fission \citep{Panov10}, and $\beta$-delayed fission including $\beta$-delayed neutron emission \citep{Mumpower2022}, with fission fragment distributions taken from \cite{Mumpower2020}. 
Temperatures are self-consistently calculated by updating the entropy with the energy released in nuclear reactions, see \cite{Freiburghaus1999} and \cite{reichert23} for more details. Neutrino energy losses from weak interactions are included, either as provided in the theoretical weak rate tables or, for Reaclib rates, based on experimental measurements from the ENSDF database \citep{Brown2018}. For $\beta$-decays without available experimental or theoretical neutrino energy data, we assume that 35\% of the nuclear energy is lost via neutrino emission \citep{reichert23,wollaeger18a}. 

\section{Parametrized Trajectories}
\label{sec:trajectories}
\begin{figure*}
  \centering
  \includegraphics[width=\textwidth]{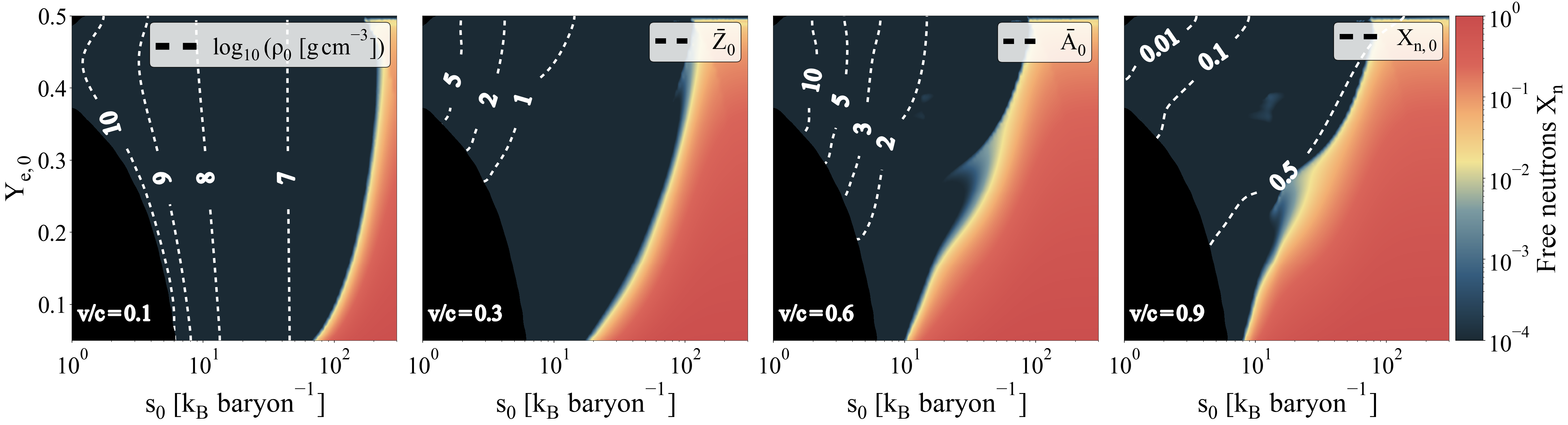}
  \caption{Mass fractions in free neutrons at $1 \,\text{min}$ after the merger across the parameter space for a constant initial temperature of $T_0=10\,\rm{GK}$. The white contour lines indicate properties at the moment of ejection, including the initial density $\rho_0$, the average mass number $\bar{A}_0$, the average charge number $\bar{Z}_0$ and the initial neutron mass fraction $X_{n,0}$. Black regions correspond to $\rho_0>10^{11}\,\rm{g\,cm^{-3}}$, where the EOS is no longer valid.}
  \label{fig:initial_densities}
\end{figure*}

The analysis of the r-process in astrophysical outflows by exploring the underlying parameter space has a long history, see e.g.\ \citet{freiburghaus99a, Lippuner2015,Kuske2025}, or \cite{Cowan21} for a comprehensive review.  While these works mainly focused on the final nucleosynthesis yields under different physical conditions, and sometimes on the radioactive heating from decaying r-process nuclei, we concentrate here explicitly on the following questions: "Under which conditions can free neutrons avoid being captured?" and "What is their impact on kilonova light curves and the detection prospects for the \textit{ULTRASAT} satellite?" \\
To this end, we perform network calculations for a large set of parametrized trajectories meant to represent various outflow components from neutron star mergers. For each ejecta trajectory, we assume a spherically symmetric outflow characterized by initial specific entropy $s_0$, the initial electron fraction $Y_{e,0}$, and the characteristic expansion velocity $\vel$. \\
We adopt constant expansion velocities in our models, as any initial acceleration phase is expected to occur on timescales much shorter than those relevant for neutron captures during the r-process. Consequently, the bulk of the dynamically ejected material is unlikely to be significantly affected by an early acceleration phase. For the density evolution we use \citep[e.g.][]{Freiburghaus1997}:
\begin{equation}
    \rho(t) = \rho_0 \biggl(\frac{r_0}{r_0 + \vel t}\biggr)^3\,,
    \label{eq:rho_homologous}
\end{equation}
where the ejection radius is set to $r_0 = 100\,\rm km$ to represent the characteristic scale of two merging neutron stars, but the exact choice has very little impact on the results. The radius evolves according to $r(t) = r_0 + \vel t$, and the electron fraction and temperature are evolved self-consistently within the nuclear reaction network.\\
We determine the starting densities $\rho_0$ by using the Timmes equation of state\footnote{\url{https://cococubed.com/code_pages/eos.shtml}} (EOS) \citep{Timmes1999}, based on the respective initial temperature, specific entropy, and an ejecta composition characterized by the average mass and charge numbers, $\bar{A}$ and $\bar{Z}$. The composition is calculated from the nuclear Saha equation under conditions of nuclear statistical equilibrium (e.g.\, \cite{Cowan21, reichert23}) using the FRDM mass model. The same EOS is used in \texttt{WinNet}, which ensures consistency between the thermodynamic initialization and the subsequent nucleosynthesis calculations.  \noindent
For  densities exceeding $10^{11}\,\rm{g\,cm^{-3}}$, the physical
assumptions underlying the Timmes EOS are not fulfilled, therefore we 
exclude parameter combinations where the starting density $\rho_0$ exceeds this threshold value. In addition, at such high densities, $\beta$-decays are Pauli-blocked, which suppresses the r-process, as the reaction flow stalls without $\beta$-decays \citep{Freiburghaus1999}. This restriction removes only a small region of the low-entropy parameter
space, which, as we discuss further below, does not affect any of the results 
of this work.\\
In the subsequent analysis, we explore a sample of 40 000 ejecta trajectories. We vary the initial entropy from $1\,\rm {k_B\,baryon}^{-1}$ to $300\,\rm {k_B\,baryon}^{-1}$, the initial electron fraction from 0.05 to 0.5, and the expansion velocity from $0.1\,c$ to $0.9\,c$. We fix the initial temperature to $T_0 = 10\,\mathrm{GK}$, as the bulk of the neutron-rich ejecta are expected to reach temperatures in or close to nuclear statistical equilibrium (NSE). Even initially cold tidal ejecta may be rapidly reheated if free protons are initially present, as nucleons recombine into $\alpha$-particles. This process is described in Appendix~\ref{sec:appendix_A}, where we update a similar calculation from \cite{Freiburghaus1999}. Here we also discuss the sensitivity to the starting temperature $T_0$.\\
The initial densities $\rho_0$, compositions (in terms of the average mass number $\bar{A}_0=\sum_A A\,Y_{A,Z}/\sum_AY_{A,Z}=1/\sum_AY_{A,Z}$ and charge number $\bar{Z}_0=\sum_A Z\,Y_{A,Z}/\sum_AY_{A,Z}$), and neutron mass fractions $X_{n,0}$ under NSE conditions are shown in Figure~\ref{fig:initial_densities} as white contour lines across the parameter grid.

\section{Neutron star merger simulations}
\label{sec:NSNS_simulations}
To round up our study, we also analyze a set of three neutron-star merger simulations that were
performed with the Lagrangian Numerical Relativity code \texttt{SPHINCS\_BSSN}
\citep{rosswog21a,diener22a,rosswog22b,rosswog23a,rosswog25c}. 
All simulations employ 2 million SPH particles and use
a 
piecewise-polytropic version \citep{read09} of the ENG equation of state \citep{engvik96} for the cold part which is consistent with known constraints \citep{biswas22b}. We include thermal contributions to internal energy and pressure based on Landau's Fermi liquid theory \citep{baym04,constantinou15} as suggested by \cite{raithel19,raithel21a}, for the corresponding implementation into the \texttt{SPHINCS\_BSSN} code see 
\citet{Biswas26}. As usual, we define the merger time as the moment 
of the first strong compression, where the lapse $\alpha$ function reaches a minimum. 
An overview of all simulations and key ejecta properties are provided in 
Table~\ref{tab:simulation-overview}.\\
We identify the ejecta at the final simulation snapshot as the particles that
simultaneously satisfy a) the Bernoulli criterion \citep[e.g.][]{rezzolla13a,foucart21b}, $\Gamma_\infty\equiv - \mathcal{E}u_t > 1$,
where $\mathcal{E}$ is the specific enthalpy, b) move
outward $\vec{r} \cdot \vec{\vel} > 0$, and c) have travelled a minimum distance of 100 code units (${\approx} 150\,\mathrm{km}$) from the origin. We use the asymptotic velocity at infinity, given by $\vel/c=\sqrt{1-1/\Gamma_\infty^2}$, as the homologous expansion velocity. Due to the Lagrangian nature of the code, we can track 
the full thermodynamic history of each ejecta particle and initialize 
the network calculations once the density has dropped below 
$10^{11}\,\mathrm{g\,cm^{-3}}$ and the temperature below $10\,\mathrm{GK}$, 
consistent with the parametrized trajectories (Sec.~\ref{sec:trajectories}).\\
Since \texttt{SPHINCS\_BSSN} does currently not include neutrino transport, 
we cannot evolve the electron fraction self-consistently within the simulation. 
We therefore consider two complementary prescriptions for 
$Y_e$. In the first, we assign $Y_e$ based on the polar-angle fit of 
\citet{Setzer23}, which is obtained from 
mass-weighted angular profiles extracted from neutrino-radiation-hydrodynamics 
simulations \citep{radice18a} and accounts for the neutrino-driven 
$Y_e$ enhancement at different latitudes. 
Since these fits are dominated by the results for the (slow) bulk of
the ejecta and the fast ejecta that we are exploring here
may have a much shorter exposure to neutrinos. We therefore consider the \cite{Setzer23} prescription as an upper limit for the electron fractions in the fast ejecta. As a lower limit, we assign in a second approach to each particle the $\beta$-equilibrium value 
($Y_e \sim 0.05$; e.g.\ \cite{Farouqi22}) it carried inside the 
isolated neutron star prior to merger, 
i.e. we assume that in the fast ejecta the electron fraction has not been significantly raised compared to the pristine value inside the neutron star. 

\begin{table}
\caption{Overview of the analyzed simulations, together with key characteristics of the ejecta. The run label indicates the neutron star masses in solar units. Values for the free neutron mass $M_n$ are given for the polar-angle--dependent prescription \citep{Setzer23} of the initial $Y_e$; the corresponding values when starting from a cold $\beta$-equilibrium $Y_e$ are shown in parentheses.}
\label{tab:simulation-overview}
\centering
\footnotesize
\setlength{\tabcolsep}{5pt}
\renewcommand{\arraystretch}{1.15}
\begin{tabular}{@{}lcccccc@{}}
\toprule
\textbf{run}
& \makecell{$M_{\rm ej}$ \\ \small $[10^{-3}\,\rm M_\odot]$}
& \makecell{$M_{\rm >0.6\,\rm c}$ \\ \small $[10^{-5}\,\rm M_\odot]$}
& \makecell{$M_{\rm n}(1\,\rm min)$\\ \small $[10^{-6}\,\rm M_\odot]$}
& \makecell{$\vel_{\rm max}$ \\ \small $[c]$}
& \makecell{$\langle \vel \rangle$ \\ \small $[c]$} \\
\midrule
\texttt{2x1.2} & 2.41 & 0.00 & 1.39 (7.80) &  0.56 &  0.15 \\
\texttt{2x1.3} & 9.07 & 1.76 & 1.50 (21.60) & 0.73 & 0.19 \\
\texttt{2x1.4} & 9.63 & 3.55 & 7.36 (36.17) & 0.76 & 0.19 \\
\bottomrule
\end{tabular}
\end{table}

\section{A semi-analytical kilonovae model}
\label{sec:kn_model}

If a sufficient amount of free neutrons can avoid being captured by nuclei,
their decay can lead to a blue transient preceding the main kilonova event.
To predict corresponding precursor light curves, we employ the 1D kilonova toy model of \citet{metzger20}. We retain its core assumptions --- spherical symmetry, grey opacities, and blackbody emission --- but replace analytic heating prescriptions with rates computed using the nuclear reaction network described in Sec.~\ref{sec:network}. This approach preserves the transparency and computational efficiency of the toy model while enabling us to isolate how the early emission depends on the fastest ejecta layers and on the presence of surviving free neutrons.\\
Rather than modelling the entire ejecta, we focus exclusively on the outermost, fastest-moving layer. This choice is motivated by two key results from our calculations. First, the mass fraction of free neutrons generally increases with velocity, while the opacity decreases due to the suppression of lanthanide and actinide formation. Second, under the assumption of homologous expansion ($r \propto \vel t$), the fastest ejecta become transparent earliest. As a result, photons can escape while the ejecta are still at high temperatures (or short wavelengths) and result in  an early UV precursor.\\
We decompose a general ejecta profile $M(\vel)$ into a set of spherically symmetric shells $j$, each characterized by its homologous expansion velocity $\vel_j$, so that the total ejected mass is given by $M_{\rm ej} = \sum_j M_j$. For each shell we compute the nucleosynthesis. We adopt grey opacities and prescribe them as a function of the lanthanide and actinide mass fraction. 
This is motivated by radiative-transfer calculations which show that even trace amounts of these elements can dominate the opacity in r-process--enriched ejecta \citep[e.g.][]{kasen13a,tanaka13a}.
Concretely,  we assign each shell a time-dependent opacity $\kappa_j(t)$ by taking the Planck-mean value from \citet{tanaka20a} (their Table~1) that corresponds to the combined mass fraction $X_{{\rm lan},j}(t) + X_{{\rm act},j}(t)$.\\
The thermal evolution of a shell is computed by balancing adiabatic cooling, radioactive heating, and diffusive radiative losses, closely following \citet{metzger20}. We evolve the radiation-dominated specific internal energy $e_j(t)$ according to
\begin{equation}
\frac{{\rm d}e_j}{{\rm d}t}
= -\frac{\vel_j}{r_j}\,e_j
-\frac{e_j}{t_{{\rm diff},j}+t_{{\rm lc},j}}
+ \dot{q}^{\,\rm th}_j\,,
\label{eq:energy_ode}
\end{equation}
where $r_j = r_0 + \vel_j t$ is the shell radius. The diffusive energy flux through a layer $j$ can be written as $F_j = (c/3\kappa_j\rho_j)\,\partial_r(e_j\rho_j)$. In a homologously (and spherically symmetric) expanding outflow, the radial gradient is determined by the velocity structure of the ejecta profile, $\partial_r(e_j\rho_j) \approx e_j\rho_j/\Delta r_j$, where the characteristic photon escape scale $\Delta r_j$ follows the mass distribution (cf. \cite{metzger15a}),
\begin{equation}
    \Delta r_j \simeq \vel_j\,t\left|\frac{\partial \ln M(\geq \vel_j)}{\partial \ln \vel_j}\right|^{-1}\,,
    \label{eq:delta_r}
\end{equation}
with $M(\geq \vel_j)$ being the cumulative mass in the exterior of shell $j$ (including its own mass $M_j$). The corresponding diffusion time is defined as
\begin{equation}
    t_{{\rm diff},j} \simeq \frac{\Delta r_j}{c}\,\tau_j\,,
    \label{eq:t_diff}
\end{equation}
with the optical depth
\begin{equation}
    \tau_j= \sum_{i\geq j}\frac{M_i\,\kappa_i}{4\pi r_i^2}\,.
\end{equation}
Once the layer $j$ becomes optically thin (meaning $\tau_j\ll1$), photons cannot escape faster than on the light-crossing time
\begin{equation}
    t_{{\rm lc},j} = \frac{(\vel_{\rm max}- \vel_j)\,t}{c}\,.
    \label{eq:t_lc}
\end{equation}
The luminosity emitted by each shell is then
\begin{equation}
    L_j = \frac{e_j\,M_j}{t_{{\rm diff},j}+t_{{\rm lc},j}}\,,
\end{equation}
and the total bolometric luminosity is $L_{\rm bol}=\sum_j L_j$.\\
As initial condition for Eq.~(\ref{eq:energy_ode}), we use the approach of \citet{metzger20} and set the specific internal energy of each shell to equal its kinetic energy,
\( e_j(t_{0,j}) = \tfrac{1}{2} \vel_j^2 \).
For freely expanding ejecta, the resulting light curves are largely insensitive to this choice due to rapid adiabatic cooling.\\
A key difference between this and fully analytic models is that we take the nuclear heating rates $\dot{q}_j$ directly from our \texttt{WinNet} calculations. They already account for neutrino losses from $\beta$-decays (see Sec.~\ref{sec:network}), but they are "naked" heating rates and only a fraction of this released energy (depending on the decay products) thermalizes with
the expanding medium.\\
We divide the heating into a neutron $\dot{q}_{n,j}$ and an r-process component $\dot{q}_{{\rm rp},j} = \dot{q}_j - \dot{q}_{n,j}$, and apply separate thermalization efficiencies. For the r-process component and we adopt a density-dependent efficiency $\epsilon_{\rm th,rp}$ following \citet{wollaeger18a}, which accounts for the energy deposition of $\beta$-particles, $\gamma$-rays, and $\alpha$-particles using the termalization physics that was used in \citet{barnes16a}.\\
Since the kinetic energy of free-neutron $\beta$-decay electrons is low ($Q \approx 0.782\,\mathrm{MeV}$) and their decay occurs at relatively high densities, these electrons thermalize efficiently, see Appendix~\ref{sec:appendix_C} for further details. We therefore set $\epsilon_{{\rm th},n} = 1$. The total thermalized heating rate becomes
\begin{equation}
\dot{q}^{\,\rm th}_j = \epsilon_{\rm th,rp}(t)\,\dot{q}_{{\rm rp},j} +
\dot{q}_{n,j}\,.
\label{eq:thermalization}
\end{equation}
\\
We assume for the resulting AB magnitudes that the radiation at this stage can 
be approximated as black body emission from a photosphere $r_{\rm ph}(t)$, with effective temperature
\begin{equation}
T_{\rm{eff}} = \left[\frac{L_{\rm bol}}{4\pi \sigma_{\rm SB} 
r_{\rm ph}^2}\right]^{1/4}\,.
\label{eq:Teff}
\end{equation}
We determine the photospheric radius $r_{\rm ph}(t)$ at each time by identifying the shell $j$ where the optical depth equals $1$.\\
We compute synthetic NUV magnitudes by integrating the blackbody spectrum over the \textit{ULTRASAT} system throughput curve \citep{Ben-Ami23}\footnote{Digitized from their published figure with WebPlotDigitizer: \url{https://automeris.io/}}. We report apparent AB magnitudes at a fiducial distance of $D=200~\mathrm{Mpc}$, roughly the LIGO-Virgo-Kagra detection horizon for the beginning of O5, which can be translated to other distances via the standard scaling $m_2 = m_1 + 5 \log_{10}\left(D_2/D_1\right)$.

\section{Results}
\label{sec:results}
\subsection{Under which conditions are free neutrons produced?}
\label{sec:free_neutrons}

\begin{figure*}
  \centering
  \includegraphics[width=\textwidth]{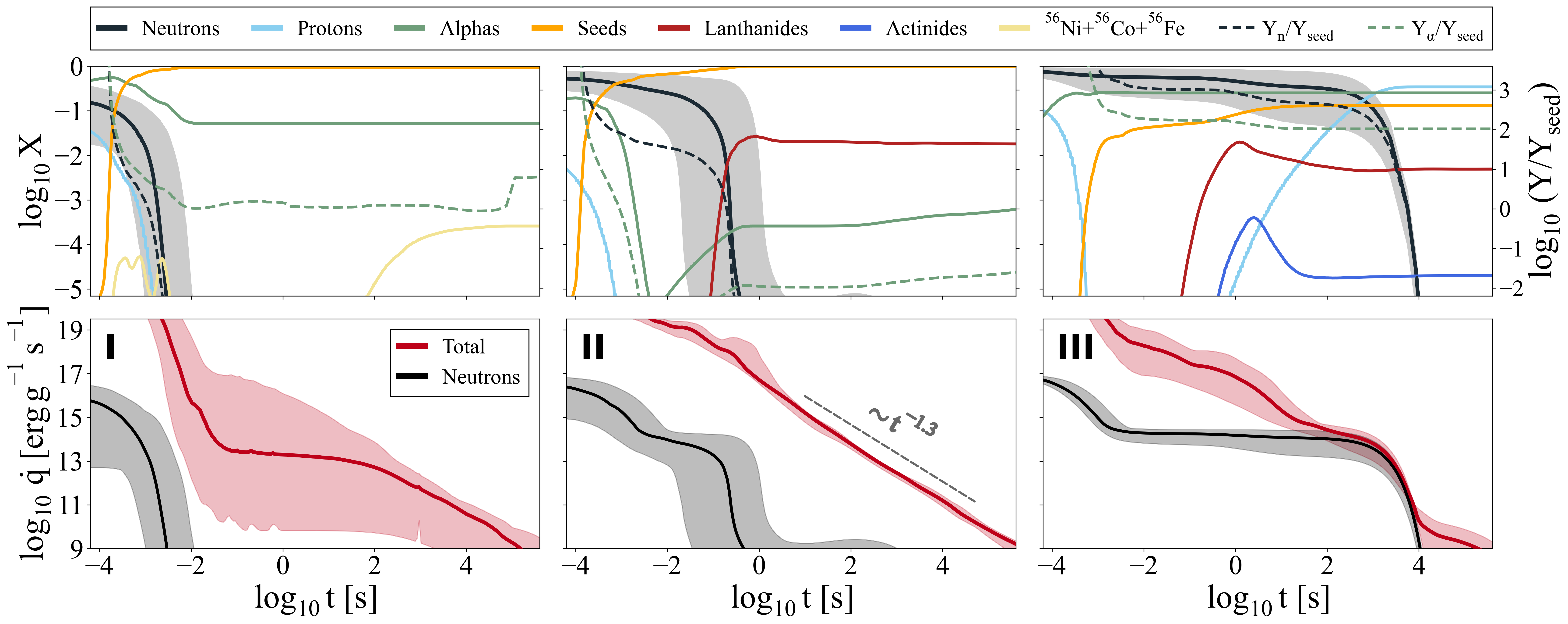}
  \caption{
  Temporal evolution of mass fractions (top) and energy generation (bottom) for the three ejecta channels (I--III). In Channel~I, neutrons are already consumed during seed formation, so no full r-process develops. In Channel~II, neutrons survive seed formation but are fully consumed during the subsequent r-process, enabling the production of heavy nuclei. In Channel~III, a substantial fraction of free neutrons survives until r-process freeze-out and their decay dominates the early heating.
  The top panels show free neutrons, free protons, $\alpha$-particles, seed nuclei ($A \ge 12$), lanthanides ($Z=57$--$71$), actinides ($Z=89$--$103$), and the iron-group elements ${}^{56}\mathrm{Ni}+{}^{56}\mathrm{Co}+{}^{56}\mathrm{Fe}$.
  Dashed curves indicate the $Y_n/Y_{\rm seed}$ and $Y_\alpha/Y_{\rm seed}$ ratios.
  Solid curves show the median, with corresponding shaded 1$\sigma$-equivalent bands (16th--84th percentiles).
  }
  \label{fig:mass_fractions_and_energy_generations}
\end{figure*}

\begin{figure*}
  \includegraphics[width=\textwidth]{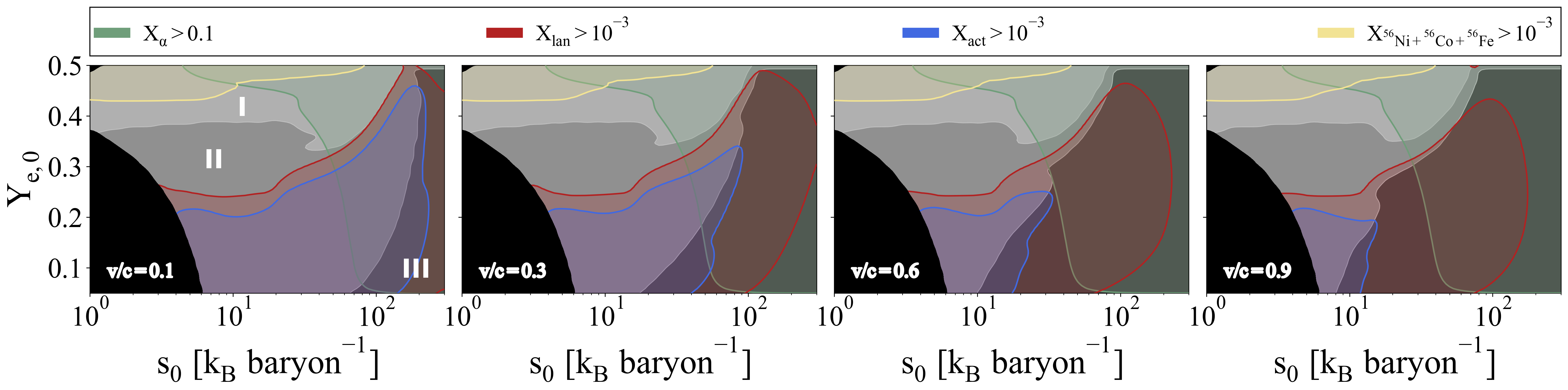}
  \caption{Parameter space divided into the three channel regions, see Table~\ref{tab:channel-definitions} and Fig.~\ref{fig:mass_fractions_and_energy_generations} (the black region indicates where the initial conditions would run out of the validity range of the Timmes EOS). We also highlight the areas where an $\alpha$-rich freeze-out occurs (green), as well as high-opacity regions where lanthanides (red) and/or actinides (blue) are produced. At high electron fractions, the ejecta form tightly bound nuclei such as Ni, Fe, and Co (yellow).}
  \label{fig:parameter_space}
\end{figure*}

We show in Figure~\ref{fig:initial_densities} the mass fraction in free neutrons within the ejecta at a time of $1\,\rm min$ after the merger when the r-process has ceased, but before a significant portion of the neutrons has decayed via $\beta^-$-decay. Each panel displays the ($Y_{e,0}$, $s_0$) plane for a different expansion velocity.\\
The parameter space separates into two distinct regimes: one in which a substantial fraction ($\gtrsim$10\%) of neutrons survives, and another one in which essentially all neutrons are captured by the end of the r-process. The transition between these two regimes is remarkably sharp and confined to a narrow band. Large entropies and expansion velocities, together with low electron fractions, favor ejecta in which neutron capture becomes inefficient, leaving a considerable fraction of free neutrons unabsorbed. While secondary processes such as $\beta$-delayed neutron emission or spontaneous fission can release neutrons again, we find that their overall contribution to the free neutron budget to be negligible.\\
For later use, we define the time of the charged-particle freeze-out $t_{\rm CPFO}$, 
indicating the end of the seed formation. We identify this moment from the combined 
conditions $T<3\,\rm GK$ and $\left|\mathrm{d}X_\alpha/\mathrm{d}t\right|<10^{-4}\,\rm{s}^{-1}$.
At this time we evaluate the neutron-to-seed ratio
$n_s(t_{\rm CPFO}) \equiv Y_n/Y_{\rm seed}$ (with $Y_{\rm seed}=\sum_{A\geq12}Y(A)$).
We further define the neutron exhaustion time $t_{\rm exh}$ as the instant when the 
neutron mass fraction drops below a small threshold, for which we use $X_n(t_{\rm exh}) < 10^{-3}$.
Finally, we define the r-process freeze-out time $t_{\rm RPFO}$ from the behavior 
of the neutron-capture reactions $(n,\gamma)$, $(n,p)$, and $(n,\alpha)$.
Once the minimum of their timescales $\tau_{n\gamma},\tau_{np}$ and $\tau_{n\alpha}$
exceeds $100\,\rm s$, the r-process is considered to have frozen out.\\
To better understand the fate of the ejected neutrons, Figure~\ref{fig:mass_fractions_and_energy_generations} shows the temporal evolution of their mass fraction.
Based on the behavior of $X_{\rm n}(t)$, the ejecta models can be broadly grouped into three channels, which correspond to distinct histories of seed formation and neutron consumption. The channels are:
\begin{itemize}
\item {\bf Channel~I} includes trajectories where neutrons run out \emph{before} charged-particle freeze-out.
We define it by $t_{\rm exh} < t_{\rm CPFO}$. In this regime, the r-process cannot proceed because too few neutrons remain at $t_{\rm CPFO}$ once seed formation is complete.
\item For the trajectories of {\bf Channel~II}, the neutrons are consumed after charged-particle freeze-out by forming heavier nuclei through a pure r-process.
\item {\bf Channel~III} is defined by a significant amount of surviving free neutrons at r-process freeze-out with $t_{\rm RPFO}<t_{\rm exh}$.
\end{itemize}
These channel definitions are summarized in Table~\ref{tab:channel-definitions}.
Regions of parameter space associated with each channel are shown as grey areas in Figure~\ref{fig:parameter_space}.\\
Channel~I occurs mainly at moderate entropies and high electron fractions $Y_{e,0}\gtrsim0.38$, whereas Channel~III occupies the high-entropy region (cf. \cite{farouqi10}).
Free neutrons also appear for moderate entropies $\sim10\,\rm k_B\,per\,baryon$ when $Y_{e,0}$ is low and the expansion is sufficiently fast. This may have interesting consequences for mergers with large amounts of tidal ejecta, as we discuss  below. Moreover, the shape of the free-neutron boundary closely follows the seed contour lines at CPFO. This is especially true for moderately slow outflows and it breaks down for high-velocities and low electron fractions, where usually more neutrons survive and the outcome becomes more sensitive to the detailed nuclear reaction path.\\ 
The boundary between the Channel~I and Channel~II region, on the other hand, does not depend notably on $v$, which is expected because the (n,$\gamma$) reactions operate on timescales much shorter than the dynamical expansion ($\tau_{\rm dyn}\approx r/\vel\sim 1\, \rm ms$).
\begin{table}
\caption{Ejecta channels defined by the temporal evolution of seed 
formation and neutron consumption. $t_{\rm exh}$ denotes the neutron 
exhaustion time, $t_{\rm CPFO}$ the charged-particle freeze-out time and $t_{\rm RPFO}$ the r-process freeze-out time.}
\label{tab:channel-definitions}
\centering
\footnotesize
\setlength{\tabcolsep}{5pt}
\renewcommand{\arraystretch}{1.15}
\begin{tabular*}{\columnwidth}{@{\extracolsep{\fill}}lll@{}}
\toprule
\textbf{Channel} & \textbf{Feature} & \textbf{Criterion} \\
\midrule
\textbf{I}   & Early neutron exhaustion &
$t_{\rm exh}<t_{\rm CPFO}$ \\
\textbf{II}  & Exhaustion during r-process &
$t_{\rm CPFO}<t_{\rm exh}<t_{\rm RPFO}$ \\
\textbf{III} & Free neutrons survive &
$t_{\rm RPFO}<t_{\rm exh}$ \\
\bottomrule
\end{tabular*}
\end{table}
%
%

The bottom row in Figure~\ref{fig:mass_fractions_and_energy_generations} shows the total specific heating rate in erg/(g s) and the neutron contribution. A power-law fit to the average total heating rate in the r-process channel (Channel II) yields a power-law exponent of $1.3$, which is considered a tell-tale  r-process signature \citep{metzger10b,korobkin12a,Lippuner2015,hotokezaka17a} and has been observed for the first neutron star merger event, see e.g.\ Fig.~1 in \cite{rosswog18a}.
In Channel~III, neutron decay provides most of the thermal energy during the first seconds to hours after the merger, whereas in the other channels its contribution to the total heating is completely negligible. We find that already a neutron mass fraction of only $X_n(1\,\mathrm{min}) \gtrsim 0.05$ is sufficient for neutron decay to dominate the total heating rate.

\subsubsection{An analytical criterion for free neutrons}
\label{sec:analytic_criterion}

\begin{figure*}
  \includegraphics[width=\textwidth]{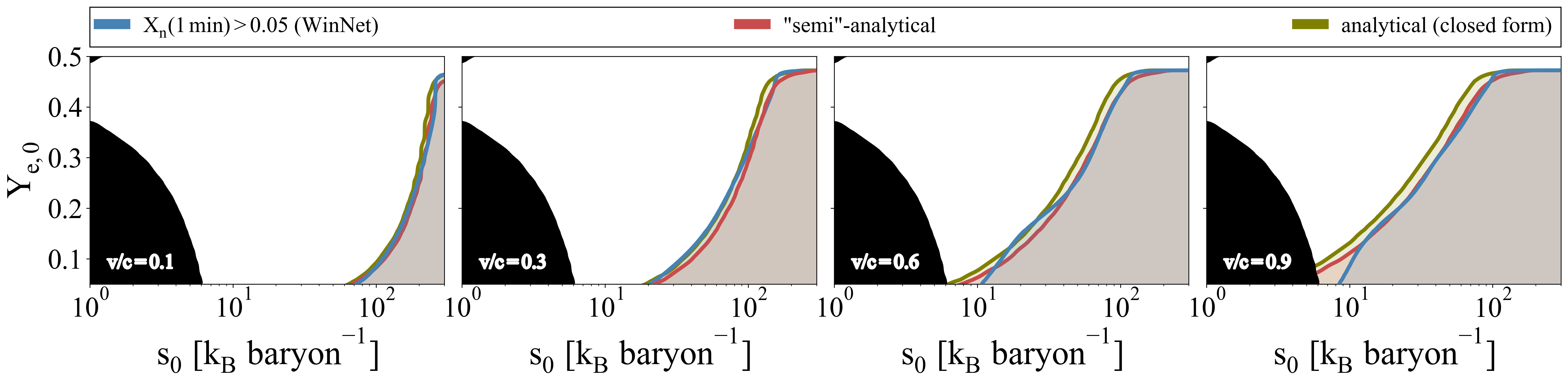}
  \caption{Regions of initial conditions yielding a free-neutron mass fraction above 5 per cent one minute after the merger. Shown are the region obtained from the \texttt{WinNet} network calculation (blue), the prediction of the semi-analytical criterion from Sect.~\ref{sec:analytic_criterion} (red), and that of the closed-form expression in Eq.~(\ref{eq:closed_form}) (green). Black areas indicate excluded high-density trajectories.}
  \label{fig:analytical_criterion}
\end{figure*}

In this section we derive a compact analytic criterion for the survival of free neutrons, i.e. for Channel~III, that is consistent with the setup of our trajectories (Sec.~\ref{sec:trajectories}). For this model, we start with a charge neutral mix of protons, electrons and neutrons, i.e. we use
 $X_{p,0}=Y_{e,0}$ and $X_{n,0}=1-Y_{e,0}$. We adopt an adiabatic temperature law that is consistent with our density evolution of 
\begin{equation}
    T(t)=T_0\,\biggl(\frac{r_0}{r_0+\vel t}\biggr)\,,
\label{eq:T_analytic}
\end{equation}
 as e.g.\ in \cite{Fujimoto08}, and we use the shorthand $T_9\equiv T/(10^9\,{\rm K})$, $\rho_5\equiv\rho/(10^5\,{\rm g\,cm^{-3}})$ and $\hat{s}_0\equiv s_0/(1\, \rm k_B \,per\, baryon)$. We further assume a constant entropy $s(t)=s_0$ that accounts for radiation and relativistic $e^\pm$-pairs (which partly recombine into photons at $T_9<5$ \citep{Hoffman97})
\begin{equation}
    \rho_5(t)\approx3\,\frac{T_9(t)^3}{\hat{s}_0}\,.
\label{eq:srad_closure_analytic}
\end{equation}
 As the material cools, nucleons recombine into $\alpha$-particles. We define the onset of the $\alpha$-process, $t_{\alpha,0}$, as the time when the temperature has dropped to $T_9(t_{\alpha,0}) = 7$, where $\alpha$-captures begin to dominate over their inverse photodisintegration rates. For neutron-rich matter ($Y_{e,0}<0.5$), we assume that by this time essentially all protons are bound into helium. This yields
\begin{equation}
    X_{n,\alpha_0} \equiv X_n(t_{\alpha,0}) = 1 - 2Y_{e,0},
    \quad
    X_{\alpha,\alpha_0} \equiv X_\alpha(t_{\alpha,0}) = 2Y_{e,0}\,,
\end{equation}
and therefore
\begin{equation}
    Y_{n,\alpha_0} \equiv X_{n,\alpha_0} = 1 - 2Y_{e,0},
    \quad
    Y_{\alpha,\alpha_0} \equiv \frac{X_{\alpha,\alpha_0}}{4} = \frac{Y_{e,0}}{2}\,.
\end{equation}
Charged-particle reactions are in this model assumed to freeze out at $T_{9}=2.5$, defining the time $t_{\rm CPFO}$. \\
For the seed formation, we follow the $\alpha$-process model of \citet{Hoffman97}, who reduce the nuclear network to the single dominant reaction chain in neutron-rich ejecta ${}^{4}{\rm He}(\alpha n,\gamma){}^{9}{\rm Be}(\alpha,n){}^{12}{\rm C}$. 
Because of its low Q-value, the bottleneck reaction ${}^{4}{\rm He}(\alpha n,\gamma){}^{9}{\rm Be}$ is expected to be in quasi-statistical equilibrium (QSE), so that at a given $T_9$
the beryllium abundance is
\begin{equation}
Y_{^{9}{\rm Be}}\approx 8.66\times 10^{-11}\,Y_\alpha^2\,Y_n\,\rho_5^{2}\,T_9^{-3}\,
\exp\!\left(\frac{18.26}{T_9}\right).
\end{equation}
Seed production is therefore driven by ${}^{9}{\rm Be}(\alpha,n){}^{12}{\rm C}$ with rate $\lambda_{\alpha n}(T)\equiv N_A\langle\sigma \vel\rangle_{\alpha n}(T)$, for which we use the fit of \citet{Wrean94}. The evolution of the helium and neutron abundances can then be approximated by
\begin{align}
    \frac{dY_\alpha}{dt}\approx-F\,Y_\alpha\,Y_{^{9}{\rm Be}}\,\rho \, \lambda_{\alpha n},\\
    \frac{dY_n}{dt}\approx-G\,Y_\alpha\,Y_{^{9}{\rm Be}}\,\rho\,\lambda_{\alpha n},
\end{align}
where $F\equiv \bar{Z}_{\rm seed}/2$ and $G\equiv \bar{A}_{\rm seed}-2\bar{Z}_{\rm seed}$ are the numbers of $\alpha$-particles and neutrons that make up a representative seed nucleus of mass number $\bar{A}_{\rm seed}$ and charge number $\bar{Z}_{\rm seed}$.\\
With our temperature evolution, it is convenient to write this coupled system in terms of $T_9$, using Eq.~(\ref{eq:T_analytic}) we obtain the transformation $dt/dT_9=-(r_0/\vel)\,(T_{9,0}/T_{9}^{2})$. 
Rewriting the ODE's for $Y_n$ and $Y_\alpha$ in terms of $T_9$ yields
\begin{align}
    \frac{{\rm d}Y_\alpha}{{\rm d}T_9}
    &=
    F\,Y_\alpha^3\,Y_n\,
    K(T_9)\,
    \frac{r_0}{\vel}\,,
    \label{eq:dYa/dT9}
    \\
    \frac{{\rm d}Y_n}{{\rm d}T_9}
    &=
    G\,Y_\alpha^3\,Y_n\,
    K(T_9)\,
    \frac{r_0}{\vel}\,,
    \label{eq:dYn/dT9}
\end{align}
    with the temperature sensitivity collected in
    \begin{equation}
    K(T_9)=
    8.66\times10^{-6}\,
    \rho_5(T_9)^3\,
    T_9^{-5}\,T_{9,0}\,
    \exp\!\left(\frac{18.26}{T_9}\right)\,
    \lambda_{\alpha n}(T_9)\,.
    \label{eq:K(T9)}
\end{equation}
Dividing the two equations leads to
\begin{equation}
    \frac{{\rm d}Y_n}{{\rm d}Y_\alpha}=\frac{G}{F}
    \quad\Rightarrow\quad
    Y_n(T_9)=Y_{n,\alpha_0}+\frac{G}{F}\left[Y_\alpha(T_9)-Y_{\alpha,\alpha_0}\right]\,,
    \label{eq:Yn_linear_in_Ya}
\end{equation}
so the coupled system reduces to a single ODE for $Y_\alpha(T_9)$,
\begin{equation}
    \frac{{\rm d}Y_\alpha}{{\rm d}T_9}
    =
    F\,Y_\alpha^3\left [
    Y_{n,\alpha_0}+\frac{G}{F}\left(Y_\alpha-Y_{\alpha,\alpha_0}\right)
    \right]\,
    K(T_9)\,
    \frac{r_0}{\vel}\,.
    \label{eq:Ya_single_ode}
\end{equation}
In practice, we integrate Eq.~(\ref{eq:Ya_single_ode}) numerically from $T_9=7$ to $T_9=2.5$, where charged-particle freeze-out is assumed to occur, while updating $Y_n$ at each step using Eq.~(\ref{eq:Yn_linear_in_Ya}).
At $t_{\rm CPFO}$, we assume that the composition consists 
exclusively of free neutrons, alpha particles 
and seed nuclei. In other words, we approximate $X_{\rm seed}\approx1-X_{n}-X_{\alpha}$,
so that the seed abundance can be written as $Y_{\rm seed}=X_{\rm seed}/\bar A_{\rm seed}$. Unless stated otherwise, we take $\bar A_{\rm seed}=90$ and $\bar Z_{\rm seed}=36$ for a typical seed distribution at CPFO.\\
After CPFO, we assume that seed formation has come to an end, i.e. $Y_{\rm seed}(t>t_{\rm CPFO})\approx Y_{\rm seed, \rm CPFO}$, and we neglect photodisintegration reactions, in particular $(\gamma,n)$, due to the low temperatures. The subsequent cold r-process can therefore be simplified to
\begin{align}
    \frac{dY_n}{dt}
    &\approx 
    \underbrace{- \rho\,Y_{\rm seed}\,Y_n\,\lambda^{\rm eff}_{n\gamma}}_{\text{captures}}
    \;-\;
    \underbrace{\frac{Y_n}{\tau_n}}_{\text{decay}}
    \label{eq:dY_n(t)},
\end{align}
where $\lambda^{\rm eff}_{n\gamma}\equiv\sum_i Y_i\,N_A\langle \sigma \vel\rangle_{i,n\gamma}/Y_{\rm seed, CPFO}$ is an effective neutron capture rate after CPFO. As neutrons do not have to overcome Coulomb barriers, we assume that $\lambda^{\rm eff}_{n,\gamma}$ does not depend on the temperature. Since $\lambda^{\rm eff}_{n,\gamma}$ varies by many orders of magnitude across the nuclear chart (from very large values near the valley of $\beta$-stability to effectively zero close to the neutron drip line) it cannot be derived within the analytic $\alpha$-process model, that does not predict the detailed seed distribution at CPFO and therefore cannot determine the exact r-process path. However, the fate of the neutrons and their dependence on the initial parameters is already largely encoded in $Y_{n,\rm CPFO}$ and $Y_{\rm seed,\rm CPFO}$ through $Y_{e,0}$, as well as in the density evolution set by $\vel$ and $s_0$. The physics not modeled in this simplified framework is then effectively absorbed into $\lambda^{\rm eff}_{n,\gamma}$. Based on our parametrized trajectories, we find that the effective neutron-capture rate along the boundary separating the free-neutron regime from the neutron-exhausted regime can be well described by 
\begin{equation}
    \lambda^{\rm eff}_{n,\gamma}(\vel,Y_{e,0})
    \approx
    \lambda^{\rm eff}_0\, \left( \frac
    {\vel}{0.3\,c} \right)^{-\eta}\,,
\end{equation}
with the neutron excess $\eta=1-2Y_{e,0}$ and $\lambda^{\rm eff}_0=6.1638\,\rm cm^3\,g^{-1}\,s^{-1}$. If we further express Eq.~(\ref{eq:dY_n(t)}) in terms of temperatures and again use Eq.~(\ref{eq:srad_closure_analytic}), we get
\begin{align}
    \frac{dY_n}{dT_9}&\approx \rho\,Y_{\rm seed}\,Y_n\,\lambda^{\rm eff}_{n,\gamma}\,\frac{r_0}{\vel}\,\frac{T_{9,0}}{T_9^2}\,+\,\frac{Y_n}{\tau_n}\frac{r_0}{\vel}\frac{T_{9,0}}{T_9^2}\,.
    \label{eq:dY_n(T_9)}
\end{align}
Integrating Eq.~(\ref{eq:dY_n(T_9)}) from CPFO to a temperature $T_9(t)$ gives 
\begin{align}
    Y_n(t)=Y_{n,\rm CPFO}\,e^{-\Theta_{n\gamma}(t)}\,,
    \label{eq:25}
\end{align}
where we introduced an effective ``neutron-capture optical depth''
\begin{equation}
    \begin{split}
    \Theta_{n\gamma}(t) &\equiv Y_{\rm seed, \rm CPFO}\,\lambda^{\rm eff}_{n,\gamma}\,T_{9,0}\,\frac{r_0}{\vel}
    \int^{T_{9,\rm CPFO}}_{T_9(t)} \frac{\rho(T_9)}{T_9^2} \, dT_9 \\
    &+ \frac{r_0\,T_{9,0}}{\tau_n\,\vel}
    \int^{T_{9,\rm CPFO}}_{T_9(t)} \frac{1}{T_9^2} \, dT_9
    \end{split}
    \label{eq:Theta_T9}
\end{equation}
At late times $T_9(t\gg t_{\rm CPFO})\approx0$ this simplifies to 
\begin{equation}
    \begin{split}
    \Theta_{n\gamma}(t) &\approx 1.5\times10^5\,Y_{\rm seed,CPFO}\,\lambda^{\rm eff}_{n,\gamma}\,T_{9,0}\,T_{9,\rm CPFO}^2\,\frac{r_0}{\hat{s}_0\,\vel}+\frac{t}{\tau_n}\,.
    \end{split}
    \label{eq:Theta_T9_approx}
\end{equation}
Using a neutron mass fraction threshold $X_n(t= 1\,\rm min) >0.05$
after the r-process has ceased (above which neutron decay dominates the total heating), we obtain a criterion for the free-neutron regime that depends only on the initial parameters $\vel$, $s_0$, $Y_{e,0}$ (and $T_0$). Within our parameter grid, this model predicts about $98\,\%$ of all points correctly.\\
Note that, under the additional assumptions that the free-neutron fraction remains unchanged during seed formation, $X_{n,\rm CPFO}\approx X_{n,\alpha_0}$, and that only $\alpha$-particles are consumed, the problem simplifies further. We evaluate the integral over the temperature-dependent part obtained from integrating Eq.~(\ref{eq:Ya_single_ode}) once numerically. This yields
\begin{equation}
    \int_{T_{9,\rm CPFO}}^{T_{9,\alpha_0}} K(T_9)\,dT_9
    \approx
    \frac{4.527\times10^9}{\hat{s}_0^3}\,\rm cm^3\,g^{-1}\,s^{-1}.
\end{equation}
The condition $X_n(t= 1\,\rm min) >0.05$ can now be expressed in fully analytical closed form as
\begin{equation}
\begin{split}
    1 > \frac{1}{\ln\Bigl(\frac{1-2Y_{e,0}}{0.05}\Bigr)}
    \Biggl\{ \frac{428.327}{\hat{s}_0}\,Y_{e,0}\,0.3^{1-2Y_{e,0}}\,
    \hat{\vel}^{2Y_{e,0}-2} \\
    \times \Biggl[ 1 - \Biggl(1 + \frac{1.36\times10^7}{\hat{s}_0^3\,\hat{\vel}}
    \bigl(Y_{e,0}^2 - 2Y_{e,0}^3\bigr)\Biggr)^{-1/2} \Biggr]
    + 0.068 \Biggr\}
    \end{split}
    \label{eq:closed_form}
\end{equation}
where we introduced $\hat{\vel}=\vel/c$.
A comparison between the analytical and semi-analytical prediction and the full nuclear reaction network values for the free-neutron boundary is shown in Fig.~\ref{fig:analytical_criterion}.\\ 
The analytical criterion slightly overestimates the surviving free-neutron fraction and shifts the predicted boundary toward lower entropies.

\subsubsection{Nucleosynthesis in the three channels}
Turning now to the nucleosynthetic outcomes of the three channels, we find that in {\bf Channel~I} the ejecta never reach a neutron-to-seed ratio high enough to trigger a full r-process. Neutrons are mainly consumed during seed formation, which leaves behind mostly iron-group nuclei such as $^{56}$Ni, $^{56}$Co, and $^{56}$Fe for $Y_{e,0}\gtrsim0.45$. At moderately high electron fractions between $0.37$ and $0.45$, however, the neutron exposure is sufficient to push material to the first r-process peak (cf. \citet{Kuske2025}).\\
{\bf Channel~II} reaches significantly higher neutron-to-seed ratios. This allows successive neutron captures to proceed into the heavy-element regime and leads to the production of lanthanides and actinides. As shown in Figure~\ref{fig:mass_fractions_and_energy_generations}, these heavy nuclei form early, before free neutron decay becomes important. Their contribution to the opacity is therefore already set by the time neutron decay becomes 
relevant, and cannot be neglected even in the fast ejecta.\\
{\bf Channel~III} tells a different story. Seed formation through the $\alpha$-process chain ${}^4\mathrm{He}(\alpha n,\gamma){}^9\mathrm{Be}(\alpha,n){}^{12}\mathrm{C}$ depends strongly on density $\dot{Y}_{\rm seed}\propto \rho^3$, see Eq.~(\ref{eq:K(T9)}) and Eq.~(\ref{eq:Ya_single_ode}). Because $\rho \propto T^3/s_0$, high entropies correspond to low densities at fixed temperature and seed 
formation stalls well before the charged-particle phase is complete. As a result, we find the ejecta freezing out with high neutron-to-seed ratios and substantial $\alpha$-mass fractions ($\alpha$-rich freeze-out). The few seeds that do form can, however, capture  enough of the abundant neutrons to be driven all the way up to the heaviest elements, so lanthanides and actinides are not entirely absent 
from Channel~III. The bulk of the neutrons, though, never find a capture target and therefore survive the r-process.

\subsection{Results for the numerical relativity trajectories}
\label{sec:sim_results}
\begin{figure*}
  \includegraphics[width=\textwidth]{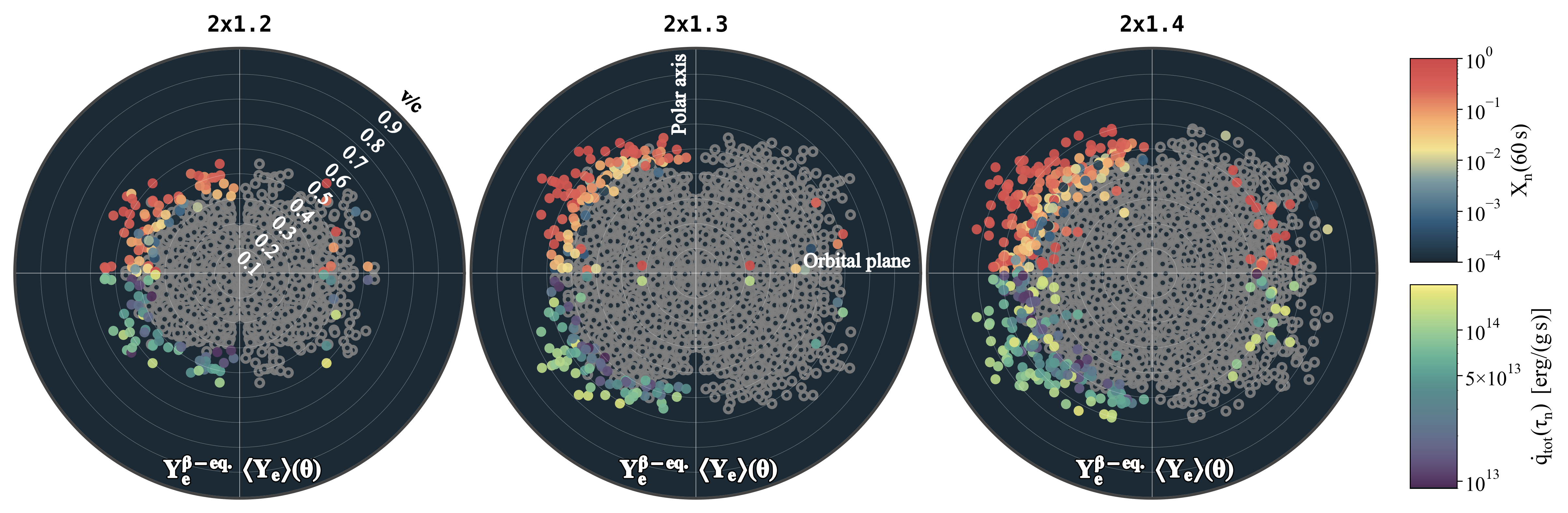}
  \caption{Free neutron mass fraction and total energy generation rate in the dynamical ejecta of three neutron star merger simulations. Each unbound SPH particle is plotted as a function of its velocity coordinate $\vel/c$ and the angle between its velocity vector and the orbital plane at the final simulation snapshot. This choice of coordinates (motivated by rotational symmetry) allows us to represent the entire ejecta of a simulation within one quarter of a circle. Each panel is therefore devided into: the free neutron mass fraction $X_n$ at $t = 1\,\mathrm{min}$ post-merger (top) and the total nuclear energy generation rate $\dot{q}_\mathrm{tot}$ at  $t =\tau_n \approx 14.6\,\mathrm{min}$ after the merger (bottom). We post-processed the trajectories using the cold $\beta$-equilibrium value $Y_e^{\beta-eq.}$ for the initial electron fraction (left), or the polar-angle-dependent fit $\langle Y_e\rangle(\theta)$ following \citet{Setzer23} (right).}
  \label{fig:simulation_results}
\end{figure*}

While the main focus of this study is the exploration of a broad range of conditions by means of parametrized trajectories, we also make use of 
ejecta histories from the three numerical relativity simulations summarized in Tab.~\ref{tab:simulation-overview}. 
As we had argued at the end of Sec.~\ref{sec:NSNS_simulations}, we consider
the electron fraction values according to
\citet{Setzer23} prescription as upper limit for the fast ejecta. Taken at face value, these $Y_e$ values 
yield hardly any free neutrons, which are restricted to a handful of
trajectories close to the orbital plane. If instead we use the lower
original $\beta$-equilibrium values, we find substantial amounts of free neutrons,  
distributed in a rather spherical way, see Fig.~\ref{fig:simulation_results}. This leading neutron shell is more pronounced for the more massive mergers which have the tendency to produce more fast ejecta. Clearly,
the true amount of free neutrons depends on the exact 
$Y_e$ value and therefore on the details of weak interactions/neutrino transport in the small amounts of mass in the outermost ejecta layers. Resolving these processes is a major challenge even for state of the art neutrino transport models.\\
In addition to the free neutrons, the lower half in the panels of Fig.~\ref{fig:simulation_results} displays the "naked" heating rates $\dot{q}$ at $t=\tau_n \approx 878\,\mathrm{s}$ after the merger. At this time, neutron decay provides the dominant contribution to the heating throughout the entire dynamical ejecta. Still, close to the polar axis, where $\langle Y_e \rangle(\theta) \gtrsim 0.35$, we also find weaker heating driven by the decay of light r-process elements, see also Sec.~\ref{sec:kn_lcs}.

\subsection{Precursor light curves}
\label{sec:kn_lcs}

\begin{figure}
    \centering
    \includegraphics[width=0.99\columnwidth]{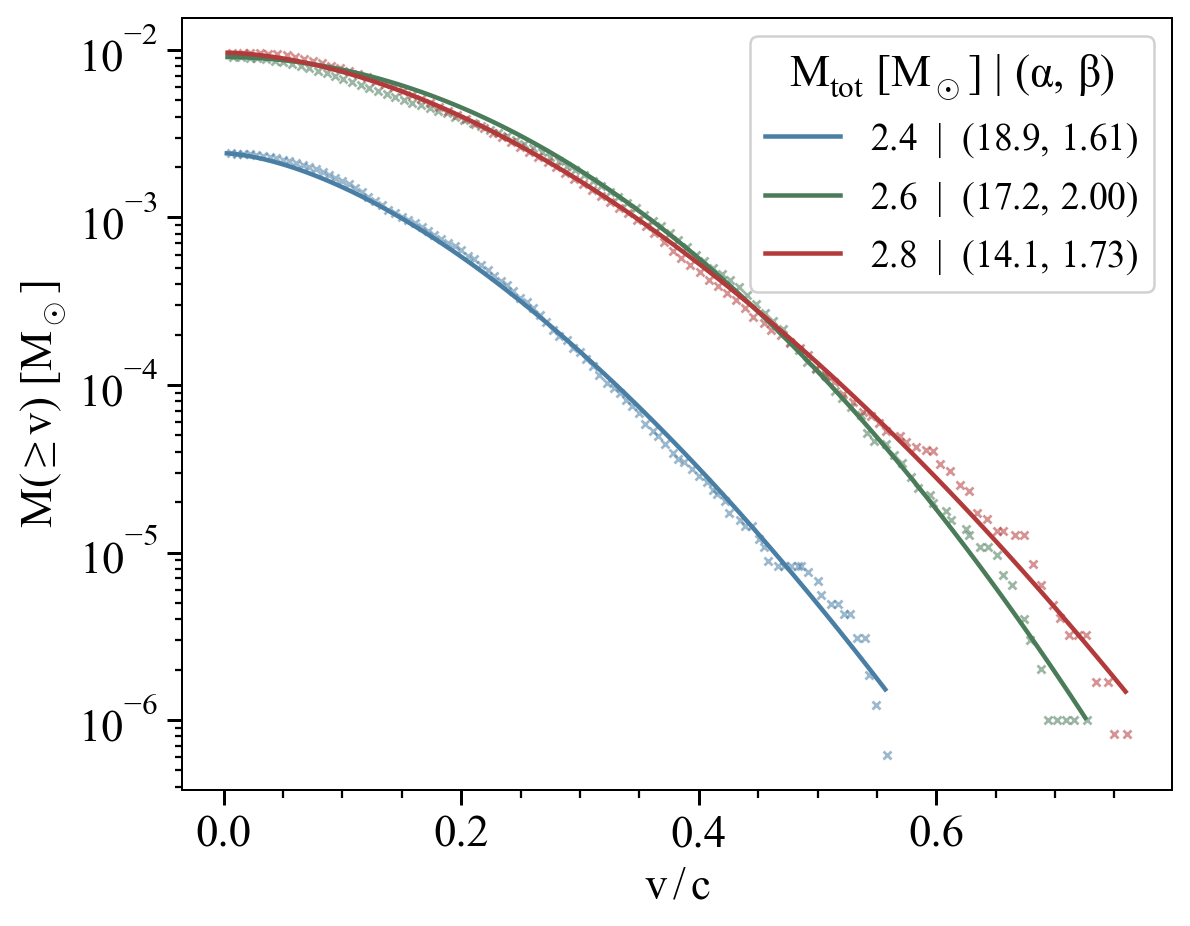}
    \caption{Cumulative ejecta mass $M(\geq \vel)$ as a function of velocity for the three
    equal-mass merger simulations (see Table~\ref{tab:simulation-overview}), together with their fits of the
    form Eq.~(\ref{eq:Mv_fit}).
    }
    \label{fig:Mv_fit}
\end{figure}

\begin{figure*}
    \centering
    \includegraphics[width=\textwidth]{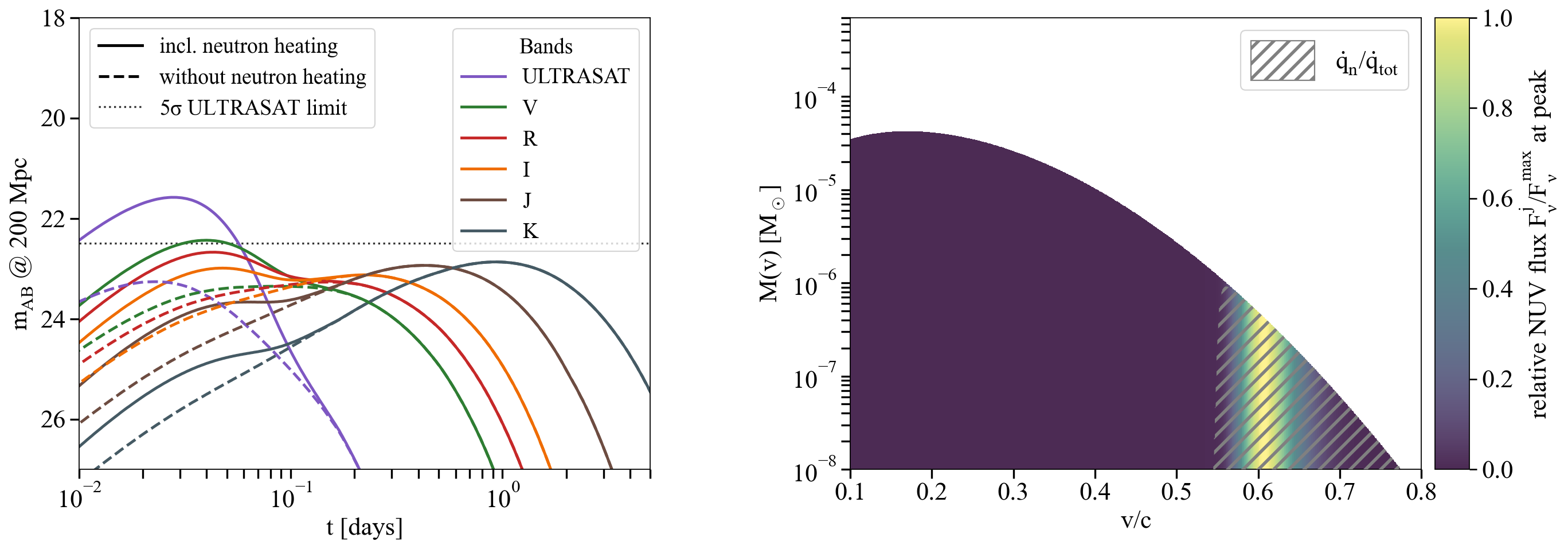}
    \caption{Kilonova light curves and mass per velocity bin for a NSM, with total binary mass $M_{\rm tot}=2.6\,\rm M_\odot$ and mass ratio $q=1$. The merger produces dynamical ejecta of $M_{\rm ej}=0.01\,\rm M_\odot$, with velocities between $\vel_{\rm min}=0.1\,c$ and $\vel_{\rm max}=0.8\,c$. The initial electron fraction is assumed to be $Y_{e,0}=0.15$ for all shells.
    \textit{Left:} Apparent AB magnitudes at $D=200$\,Mpc as a function 
    of time. Solid lines include free-neutron heating; dashed lines show 
    the r-process-only case. The dotted horizontal line marks the 
    $5\sigma$ \textit{ULTRASAT} single-exposure detection limit at $22.5\,\rm mag$.
    \textit{Right:} Mass–velocity distribution $M(\vel)$, with colours evaluated at the time of peak NUV emission in the \textit{ULTRASAT} band at ${\approx} 41\,\mathrm{min}$ after merger, when the transient reaches $m_{\rm AB}=21.6\,\mathrm{mag}$. Each bin is coloured by its relative NUV flux contribution, $F_\nu^j/F_\nu^{\rm max}$. The grey hatched fraction within each bin indicates the contribution of free-neutron $\beta$-decay to the total heating rate, $\dot{q}_n/\dot{q}_{\rm tot}$.}
    \label{fig:kn_combined}
\end{figure*}

\begin{figure}
  \centering
  \includegraphics[width=\columnwidth]{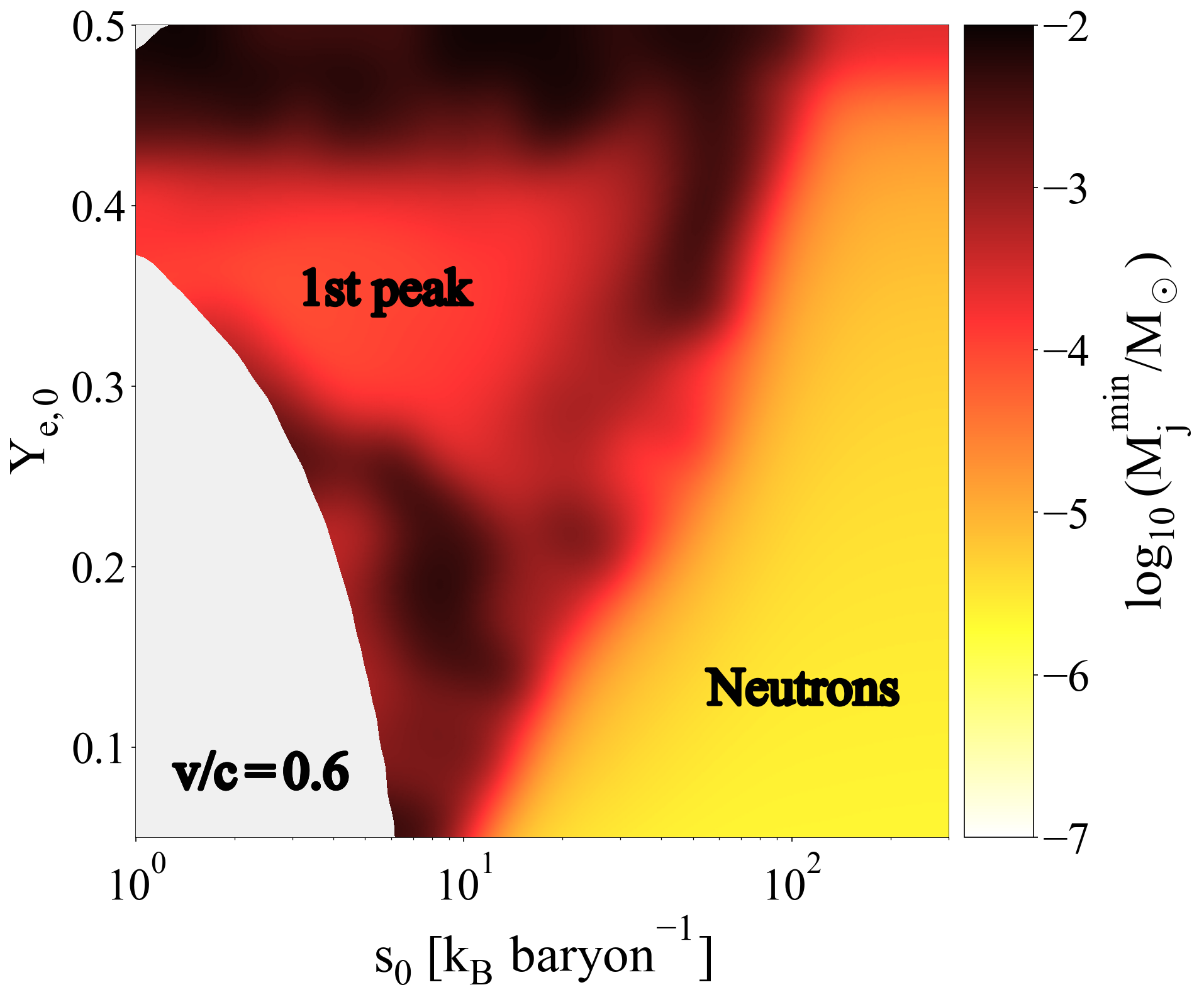}
  \caption{Minimum mass $M_j=M_{\rm min}$ in fastest ($\vel=0.6\,c$) ejecta component required for the precursor signal
  to be detectable by \textit{ULTRASAT} at a distance of $D=200\,\rm Mpc$. Here we show region, where the EOS is no longer valid in grey.}
  \label{fig:precursor_min_mass}
\end{figure}

To illustrate the model described in Sec.~\ref{sec:kn_model} and to establish a reference for our precursor calculations, we compute kilonova light curves for one representative dynamical ejecta configuration. We find that equal-mass mergers produce dynamical outflows that follow a mass distribution of the form
\begin{equation}
    M({\geq}\vel) = M_{\rm ej}\exp\bigl[-\alpha\,(\hat{\vel}^\beta - \hat{\vel}_{\rm min}^\beta)\bigr],
    \label{eq:Mv_fit}
\end{equation}
where $M({\geq}\vel)$ is the cumulative mass above velocity $\vel$, $M_{\rm ej}$ is the total ejected mass, and $\vel_{\rm min}$ is the minimum ejecta velocity (with $\hat{\vel}=\vel/c$). The ejecta are assumed to expand homologously ($\vel \propto r$), a stage which is reached soon after merger \citep{rosswog14a,neuweiler23}.\\
From our three equal-mass merger simulations (see Table~\ref{tab:simulation-overview}), the two simulations including the heavier stars, $M_{\rm tot}=2.6\,\rm M_\odot$ and $2.8\,\rm M_\odot$, undergo a similar and more violent merger. The $2.6\,\rm M_\odot$ case ejects a comparable amount of mass as the $2.8\,\rm M_\odot$ simulation but produces slightly less and slower fast ejecta. In contrast, the $2.4\,\rm M_\odot$ system shows a much weaker collision, with less ejecta and a reduced fast component, as can be seen in Fig.~\ref{fig:Mv_fit}.\\
A layer $j$ of width $\Delta \vel$ at velocity $\vel_j$ has the mass\\ $M_j\simeq|\partial M(\geq \vel_j)/\partial \vel_j|\,\Delta \vel=M(\geq \vel_j)\,\alpha\,\beta\,\vel_j^{\beta-1}\,c^{-\beta}\,\Delta \vel$ and volume $V_j\simeq4\pi (\vel t)^2\,(\Delta \vel t)$. The shell averaged density consequently is 
\begin{equation}
    \bar{\rho}_j(\vel_j,t)=\frac{M_j}{V_j}\approx \frac{\alpha\,\beta }{4\pi \,c^3\,t^3}\,\biggl(\frac{\vel_j}{c}\biggr)^{\beta-3}\,M(\geq \vel_j)\,.
\end{equation}
At the onset of the network calculation, we assume adiabatic, spherically symmetric expansion. We require the fastest (coldest) shell to be safely in NSE, which yields a temperature distribution of the form
\begin{equation}
    T(\vel_j,t) = 10\,\mathrm{GK} \left(\frac{\vel_{\max}}{\vel_j}\right) 
    \left(\frac{t}{1\,\mathrm{ms}}\right)^{-1}\,.
\end{equation}
Evaluating the photon escape scale $\Delta r_j$, see Eq.~(\ref{eq:delta_r}), for the mass profile of Eq.~(\ref{eq:Mv_fit}) leads to
\begin{equation}
    \Delta r_j = \frac{ct}{\alpha\beta}\,\biggl(\frac{\vel_j}{c}\biggr)^{1-\beta} \,,
    \label{eq:delta_rj_alpha_beta}
\end{equation}
which must be taken into account in the definition of the diffusion time, Eq.~(\ref{eq:t_diff}).\\
We apply the above model to an equal-mass neutron star merger with a total binary mass of $M_{\rm tot}=2.6\,\rm M_\odot$, where we assume a mass profile of the form of Eq.~(\ref{eq:Mv_fit}) with $\alpha=17.2$ and $\beta=2.00$. Furthermore, we assume that the dynamical ejecta has a total mass of $M_{\rm ej}=0.01\,\rm M_\odot$, with velocities ranging from $\vel_{\rm min}=0.1\,c$ to $\vel_{\rm max}=0.8\,c$ (divided into 700 shells). The resulting apparent AB magnitudes at $D=200\,\mathrm{Mpc}$ are shown in the left panel of Fig.~\ref{fig:kn_combined}. In addition to the \textit{ULTRASAT} NUV band, we show light curves computed with the standard Bessel $BVRI$ filters\footnote{SVO Filter Profile Service (Bessel filters): \url{https://svo2.cab.inta-csic.es/theory/fps/index.php?&mode=search&search_text=Generic/Bessel&zoom=1&all=0}.}. Solid lines include the contribution of free-neutron heating; dashed lines show the r-process-only case. The dotted horizontal line marks the $5\sigma$ \textit{ULTRASAT} single-exposure detection limit at $m_{\rm AB}=22.5\,\rm mag$. Neutron decay boosts the bolometric luminosity by one order of magnitude, corresponding to a difference of ${\approx} 1.7 \,\mathrm{mag}$ in apparent AB magnitude at peak emission, which occurs at ${\approx} 39\,\mathrm{min}$ after the merger.\\
The right panel of Figure~\ref{fig:kn_combined} shows where the precursor signal originates within the ejecta. It displays the binned ejecta mass $M_j$ as a function of velocity, i.e.\ the discrete representation of Eq.~(\ref{eq:Mv_fit}). The color indicates the relative contribution to the NUV flux at peak time, normalized to the maximum flux, thereby highlighting the shells responsible for the dominant emission.\\
In addition, the grey hatched regions indicate the contribution of neutron heating to the total heating at this time. The height of each bin corresponds to the total heating rate $\dot{q}_{\mathrm{tot}}$, while the hatched fraction represents the neutron heating contribution $\dot{q}_n$. We find that neutron contribution drops off sharply from nearly $100\,\%$ to essentially zero below $\vel \approx 0.56\,c$, demonstrating that the highest-velocity ejecta are required for their survival against capture during the r-process. Again, this can be linked to an ineffective seed formation and an $\alpha$-rich freeze-out in the fastest layers. Moreover, the regions with strong neutron heating coincide with those dominating the precursor flux. This shows clearly that the NUV signal is powered by neutron decay rather than by other radioactive processes. In the total outflow we saw a mass of $M_n\approx2\times10^{-5}\,\rm M_\odot$ in free neutrons at $1\,\rm min$ after the merger. \\
In the outermost parts of a steeply declining mass profile (such as the ones in Fig.~\ref{fig:Mv_fit}), the mass above a given layer $j$ is dominated by the shell itself, such that $M(\geq \vel_j) \approx M_j$. If the nucleosynthesis does not vary significantly in the overlying material, the specific heating rates are comparable, and the resulting luminosity scales as $L_j \propto M_j$, implying that it is likewise dominated by shell $j$.\\
We adopt this ''outer-layer'' approximation to estimate precursor emission from our parametrized ejecta models. Because they have constant velocities, no well-defined photosphere exists. Instead, we treat the shell itself as the only emitting surface and neglect contributions from deeper layers.\\
The interpretation is therefore conditional: if an outer shell with mass $M_j$ and parameters ($Y_{e,0}$, $s_0$, $\vel$) exists, it produces the emission discussed here. \\
In a homologous outflow the shell thickness\footnote{We set $\Delta \vel / \vel_j = 0.1$. The results shown in Fig.~\ref{fig:precursor_min_mass} depend only weakly on the precise choice of the shell thickness.} (and therefore the distance the photons have to traverse) is $\Delta r_j\simeq\bigl(\Delta \vel/\vel_j\bigr)\,r_j$. The otherwise unconstrained shell mass $M_j$ is treated as a free parameter and is sampled over the range $M_j^{\min}=10^{-7}\,\rm M_\odot$ to $M_j^{\max}=10^{-2}\,\rm M_\odot$. If the $5\sigma$ detection limit of \textit{ULTRASAT} at $200\,\rm Mpc$ ($22.5\,\rm{AB\,mag}$) falls within this range, we apply a bisection method to determine the corresponding shell mass. The minimum mass in fast ($0.6\,c$) ejecta required for \textit{ULTRASAT} detection at this distance is shown in Fig.~\ref{fig:precursor_min_mass}. For regions where the emission is driven by the decay of neutrons (i.e., channel~III), masses of a few $10^{-6}$ to $10^{-5}\,\rm M_\odot$ are sufficient for a detection.\\
In contrast, in channel~I
and~II trajectories matter accumulates at the closed
neutron shells $N\!=\!28,\,50,\,82$ and $126$ during the r-process, where the abrupt drop in neutron-capture cross sections turns each magic number into a
local waiting point. Once free neutrons are exhausted, the trapped material decays back to stability via $\beta^{-}$ chains, potentially reaching nuclei with half-lives comparable to that of free neutrons. We find that ejecta with significant material accumulated at the
first r-process peak ($0.3\lesssim Y_{e,0}\lesssim 0.4$ and $s_0\lesssim20\,\rm k_B\,per\,baryon$) show the strongest heating on the relevant
timescales, as a large fraction of this material returns to stability
through nuclei with half-lives in this range, e.g.\
$^{84}$Br ($31.8$\,min), $^{94}$Y ($18.7$\,min) or $^{90}$Rb
($2.6$\,min). This weaker heating source, however, requires a larger shell mass, so a detectable signal only emerges for ${\gtrsim} 10^{-4}\,\rm M_\odot$.\\
At this point, we cannot avoid briefly addressing the impact of special relativistic effects on the observed light curves. In general, such effects depend sensitively on the Lorentz factor $\gamma$. In our simulations, however, the outflows remain only mildly relativistic, with the fastest fluid elements reaching at most $\gamma \approx 1.54$. In addition, the sharply declining mass profiles imply that the contribution to the emission decreases rapidly toward higher velocities. The inaccuracies introduced by our Newtonian treatment are therefore expected to be subdominant compared to the broader uncertainties associated with the simplified modeling and the remaining uncertainties in the underlying nuclear physics. We therefore leave a more detailed relativistic treatment to future work.


%

%

\section{Summary and discussion}
\label{sec:summary}
In this paper we have explored the nucleosynthesis in fast dynamical ejecta of
neutron star mergers. We investigated in particular the suggestion \citep{metzger15a}
that in fast parts of these outflows free neutrons may avoid being captured and may
lead to a precursor event before the main kilonova emission. We find indeed that in a
large fraction of the ejecta parameter space, favored by large velocities, low electron fractions and high entropies, free neutrons avoid being captured by nuclei
and will lead to a ''kilonova precursor'' about one hour after merger. This indicates
that such precursors are a common feature of neutron star mergers.\\
More specifically, our main results can be summarized as follows.\\

\noindent{\em Nucleosynthesis}
\bi
\i Ejected neutrons either remain largely unbound or are almost entirely captured during the nucleosynthesis. The parameter space splits into a neutron-rich and a neutron-poor regime.
\i We find that the trajectories can be grouped in three channels, see Fig.~\ref{fig:mass_fractions_and_energy_generations} and Tab.~\ref{tab:channel-definitions}:
\bi 
\i {\em Channel I}: neutrons run out before charged particle freeze-out, so there are no neutrons left to be captured when seed formation is complete. 
\i {\em Channel II}: there are still abundant amounts of neutrons left after seed formation and they are subsequently captured by the seed nuclei in classical r-process. This channel yields the characteristic power-law heating rate $\propto t^{-1.3}$, see 
the second panel in the second row of Fig.~\ref{fig:mass_fractions_and_energy_generations}, that is considered a tell-tale for r-process \citep{metzger10b,korobkin12a,Lippuner2015} and has been observed in the bolometric light-curve decay of AT2017gfo \citep{rosswog18a}.
\i {\em Channel III}: a substantial amount free neutrons survive after the r-process has frozen out. The decay of these neutrons dominates the heating rate 
between ${\sim} 10^2$ and ${\sim} 10^4$ seconds after the merger, provided that
the free neutron mass fraction is $X_n \gtrsim 0.05$.
\ei
\i We provide (semi-)analytical criteria for the presence of a substantial amount of free neutrons in Sec.~\ref{sec:analytic_criterion}, which are inspired by the pioneering work of \citet{Hoffman97}. In its semi-analytic form, see Eq.~(\ref{eq:25}) in Sec.~\ref{sec:analytic_criterion}, the criterion correctly predicts the presence of free neutrons in ${\approx}98$ \% of the cases, and slightly less accurate in its analytic form, see Eq.~(\ref{eq:closed_form}). Analytical (green) and semi-analytical criterion (red) are shown together with the full nuclear network calculation
(blue) in Fig.~\ref{fig:analytical_criterion}.
\i For our numerical relativity simulations we find an accurate fit for the distribution of the dynamical ejecta mass (for equal-mass mergers) with velocity, see Eq.~(\ref{eq:Mv_fit}).
For the corresponding ejecta trajectories  we
hardly find free neutrons when we apply the electron fraction fit due to \citet{Setzer23}, but as discussed in Sec.~\ref{sec:NSNS_simulations}, we consider these electron fraction values as too large for the fast ejecta that are the main focus here. If instead we apply the pristine electron fraction values from the original neutron stars, we find substantial amounts of free neutrons, see Fig.~\ref{fig:simulation_results}.
\ei
{\em Free neutron decay-induced kilonova precursors}
\bi
\i Such precursors appear to be a common feature of neutron star mergers.
\i In nucleosynthesis channel~III, they dominate the early (${\sim} 1\,\rm hr$) heating rate and cause early precursor emission which we exemplify in the left panel of Fig.~\ref{fig:kn_combined}. A close inspection of this case demonstrates that the emission at peak is dominated by ejecta at $\vel\approx 0.6\,c$, see the right panel.
\i We find that a free neutron mass as small as $10^{-6}-10^{-5}\,\rm M_\odot$ suffices
to power a transient that is detectable by \textit{ULTRASAT} out to $200\,\rm Mpc$, close to the
neutron star merger detection horizon\footnote{https://observing.docs.ligo.org/plan/.} 
expected for LIGO-Virgo-Kagra in science run O5.
\i Since part of the ejecta are launched by shocks due to deep relativistic compressions of the neutron stars, see e.g.\
\cite{radice18a,combi23,rosswog25b}, we expect that these 
transients carry the imprints of nuclear matter at the highest 
densities. For example, softer equations of state lead to larger
peak velocities.
\ei

\noindent {\em Different compact binaries}\\
As we have shown, large free-neutron fractions arise particularly at low $Y_{e,0}$, i.e. in ejecta that are still close to their cold $\beta$-equilibrium values from the original neutron stars. This naturally points to cold tidal dynamical ejecta as a promising candidate for strong precursor emission, a conclusion further supported by our finding that free neutrons can survive even at moderate ejection velocities. Recent work on compact binaries with substantial mass asymmetries, either double neutron stars or neutron star black hole systems, 
has also found fast ejecta that seem to be launched by tidal torques during the merger \citep{matur24,bernuzzi25}. Such binaries with their 
large tidal ejecta masses, low electron fractions and potentially fast velocities should result in particularly bright kilonova precursor transients. \\
Neutron star binaries that are dynamically formed in locations
with large number densities such as globular clusters have a fair probability to contain a single, spun-up millisecond pulsar together with a slowly/non-spinning neutron star \citep{rosswog24b}. Such systems with large one-sided tidal tails are morphologically similar to neutron star mergers with a large mass asymmetry. With tidal tails of ${\sim} 10^{-2}\,\rm M_\odot$ and typical bulk velocities of ${\sim}0.3\,c$
they may be accompanied by bright free neutron decay-powered kilonova precursors.\\
At the same time, high specific entropies also favor free-neutron survival by suppressing seed nucleus formation. Reaching such entropies, however, requires the ejecta to be strongly shocked, as is the case for the fast ejecta produced during the collision itself \citep{radice18a,combi23,rosswog25b}. 
The electron fraction evolution in such ejecta depends on the competition between the expansion and the weak-interaction timescales. To accurately determine the physical conditions in these small amounts of ejecta is 
a serious challenge that is left for future work.

\section*{Acknowledgements}
It is a great pleasure to acknowledge stimulating discussions with Anders Jerkstrand and Blanka Világos.\\
LS and SR have been supported by  the European Research 
Council (ERC) Advanced Grant INSPIRATION under the European Union's 
Horizon 2020 research and innovation programme (Grant agreement No. 
101053985) and by Deutsche Forschungsgemeinschaft 
(DFG, German Research Foundation) under Germany's Excellence Strategy 
- EXC 2121 "Quantum Universe" - 390833306.
SR has been further supported by the Swedish Research Council (VR) under 
grant number 2020-05044, by the research environment grant
``Gravitational Radiation and Electromagnetic Astrophysical
Transients'' (GREAT) funded by the Swedish Research Council (VR) 
under Dnr 2016-06012 and  by the Knut and Alice Wallenberg Foundation
under grant Dnr. KAW 2019.0112.\\
An AI tool was used to assist in generating Fig.~\ref{fig:ejecta_overview}. The final figure was reviewed and adapted by the authors.

\section*{Data Availability}
The data underlying this article will be shared on reasonable request to the corresponding author.


\bibliographystyle{mnras}
\bibliography{astro_SKR,astro_LS} 

@article{Ben-Ami23,
  author = {Sagi Ben-Ami and Yossi Shvartzvald and Eli Waxman and Udi Netzer and Yoram Yaniv and Viktor M. Algranatti and Avishay Gal-Yam and Ofer Lapid and Eran Ofek and Jeremy Topaz and Iair Arcavi and Arooj Asif and Shlomi Azaria and Eran Bahalul and Merlin F. Barschke and Benjamin Bastian-Querner and David Berge and Vlad D. Berlea and Rolf Buehler and Louise Dittmar and Anatoly Gelman and Gianluca Giavitto and Or Guttman and Juan M. Haces Crespo and Daniel Heilbrunn and Arik Kachergincky and Nirmal Kaipachery and Marek Kowalski and Shrinivasrao R. Kulkarni and Shashank Kumar and Daniel K{\"u}sters and Tuvia Liran and Yonit Miron-Salomon and Zohar Mor and Aharon Nir and Gadi Nitzan and Sebastian Philipp and Andrea Porelli and Ilan Sagiv and Julian Schliwinski and Tuvia Sprecher and Nicola De Simone and Nir Stern and Nicholas C. Stone and Benny Trakhtenbrot and Mikhail Vasilev and Jason J. Watson and Francesco Zappon},
  title = {The scientific payload of the Ultraviolet Transient Astronomy Satellite (ULTRASAT)},
  journal = {Space Telescopes and Instrumentation 2022: Ultraviolet to Gamma Ray},
  volume = {12181},
  pages = {1218105},
  year = {2022},
  publisher = {SPIE},
  doi = {10.1117/12.2629850},
  url = {https://doi.org/10.1117/12.2629850}
}

@ARTICLE{Brown2018,
       author = {{Brown}, D.~A. and {Chadwick}, M.~B. and {Capote}, R. and
                 {Kahler}, A.~C. and {Trkov}, A. and {Herman}, M.~W. and
                 {Sonzogni}, A.~A. and {Danon}, Y. and {Carlson}, A.~D. and
                 {Dunn}, M. and {Smith}, D.~L. and {Hale}, G.~M. and
                 {Arbanas}, G. and {Arcilla}, R. and {Bates}, C.~R. and
                 {Beck}, B. and {Becker}, B. and {Brown}, F. and
                 {Casperson}, R.~J. and {Conlin}, J. and {Cullen}, D.~E. and
                 {Descalle}, M.-A. and {Firestone}, R. and {Gaines}, T. and
                 {Guber}, K.~H. and {Hawari}, A.~I. and {Holmes}, J. and
                 {Johnson}, T.~D. and {Kawano}, T. and {Kiedrowski}, B.~C. and
                 {Koning}, A.~J. and {Kopecky}, S. and {Leal}, L. and
                 {Lestone}, J.~P. and {Lubitz}, C. and
                 {M{\'a}rquez Dami{\'a}n}, J.~I. and {Mattoon}, C.~M. and
                 {McCutchan}, E.~A. and {Mughabghab}, S. and {Navratil}, P. and
                 {Neudecker}, D. and {Nobre}, G.~P.~A. and {Noguere}, G. and
                 {Paris}, M. and {Pigni}, M.~T. and {Plompen}, A.~J. and
                 {Pritychenko}, B. and {Pronyaev}, V.~G. and {Roubtsov}, D. and
                 {Rochman}, D. and {Romano}, P. and {Schillebeeckx}, P. and
                 {Simakov}, S. and {Sin}, M. and {Sirakov}, I. and
                 {Sleaford}, B. and {Sobes}, V. and {Soukhovitskii}, E.~S. and
                 {Stetcu}, I. and {Talou}, P. and {Thompson}, I. and
                 {van der Marck}, S. and {Welser-Sherrill}, L. and {Wiarda}, D. and
                 {White}, M. and {Wormald}, J.~L. and {Wright}, R.~Q. and
                 {Zerkle}, M. and {{\v{Z}}erovnik}, G. and {Zhu}, Y.},
        title = "{ENDF/B-VIII.0: The 8$^{th}$ Major Release of the Nuclear
                  Reaction Data Library with CIELO-project Cross Sections,
                  New Standards and Thermal Scattering Data}",
      journal = {Nuclear Data Sheets},
         year = 2018,
       volume = {148},
        pages = {1-142},
          doi = {10.1016/j.nds.2018.02.001},
       adsurl = {https://ui.adsabs.harvard.edu/abs/2018NDS...148....1B},
}

@ARTICLE{Cyburt2010,
       author = {{Cyburt}, Richard H. and {Amthor}, A. Matthew and {Ferguson}, Ryan and
                 {Meisel}, Zach and {Smith}, Karl and {Warren}, Scott and
                 {Heger}, Alexander and {Hoffman}, R.~D. and {Rauscher}, Thomas and
                 {Sakharuk}, Alexander and {Schatz}, Hendrik and {Thielemann}, F.~K. and
                 {Wiescher}, Michael},
        title = "{The JINA REACLIB Database: Its Recent Updates and Impact on
                  Type-I X-ray Bursts}",
      journal = {\apjs},
         year = 2010,
       volume = {189},
       number = {1},
        pages = {240-252},
          doi = {10.1088/0067-0049/189/1/240},
       adsurl = {https://ui.adsabs.harvard.edu/abs/2010ApJS..189..240C},
}

@ARTICLE{Dong2005,
       author = {{Dong}, Tiekuang and {Ren}, Zhongzhou},
        title = "{New calculations of $\alpha$-decay half-lives by the
                  Viola--Seaborg formula}",
      journal = {European Physical Journal A},
         year = 2005,
       volume = {26},
       number = {1},
        pages = {69-72},
          doi = {10.1140/epja/i2005-10142-y},
       adsurl = {https://ui.adsabs.harvard.edu/abs/2005EPJA...26...69D},
}

@ARTICLE{Freiburghaus1997,
       author = {{Freiburghaus}, Ch. and {Rauscher}, T. and {Thielemann}, F.-K. and
                 {Kratz}, K.-L. and {Pfeiffer}, B.},
        title = "{The r-process in the high entropy bubble}",
      journal = {Nuclear Physics A},
         year = 1997,
       volume = {621},
       number = {1},
        pages = {405-408},
          doi = {10.1016/S0375-9474(97)00280-7},
       adsurl = {https://ui.adsabs.harvard.edu/abs/1997NuPhA.621..405F},
}

@ARTICLE{Freiburghaus1999,
       author = {{Freiburghaus}, C. and {Rosswog}, S. and {Thielemann}, F.-K.},
        title = "{r-Process in Neutron Star Mergers}",
      journal = {\apjl},
         year = 1999,
       volume = {525},
       number = {2},
        pages = {L121},
          doi = {10.1086/312343},
       adsurl = {https://ui.adsabs.harvard.edu/abs/1999ApJ...525L.121F},
}

@article{Fujimoto08,
doi = {10.1086/529416},
url = {https://doi.org/10.1086/529416},
year = {2008},
month = {jun},
publisher = {},
volume = {680},
number = {2},
pages = {1350},
author = {Fujimoto, Shin-ichiro and Nishimura, Nobuya and Hashimoto, Masa-aki},
title = {Nucleosynthesis in Magnetically Driven Jets from Collapsars},
journal = {The Astrophysical Journal}
}

@ARTICLE{Khuyagbaatar2020,
       author = {{Khuyagbaatar}, J.},
        title = "{Spontaneous fission half-lives of the heaviest nuclei:
                  Semi-empirical predictions}",
      journal = {Nuclear Physics A},
         year = 2020,
       volume = {1002},
        pages = {121958},
          doi = {10.1016/j.nuclphysa.2020.121958},
       adsurl = {https://ui.adsabs.harvard.edu/abs/2020NuPhA100221958K},
}

@ARTICLE{Kravchuk2014,
       author = {{Kravchuk}, P.~A. and {Yakovlev}, D.~G.},
        title = "{Strong plasma screening in thermonuclear reactions:
                  Electron drop model}",
      journal = {\prc},
         year = 2014,
       volume = {89},
       number = {1},
        pages = {015802},
          doi = {10.1103/PhysRevC.89.015802},
       adsurl = {https://ui.adsabs.harvard.edu/abs/2014PhRvC..89a5802K},
}

@misc{Kuske2025,
      author        = {{Kuske}, Jan and {Arcones}, Almudena and {Reichert}, Moritz},
      title         = "{Complete survey of r-process conditions: the
                        (un-)robustness of the r-process(-es)}",
      year          = {2025},
      eprint        = {2506.00092},
      archivePrefix = {arXiv},
      primaryClass  = {astro-ph.HE},
      url           = {https://arxiv.org/abs/2506.00092},
}

@ARTICLE{Lippuner2015,
       author = {{Lippuner}, Jonas and {Roberts}, Luke F.},
        title = "{r-Process Lanthanide Production and Heating Rates in Kilonovae}",
      journal = {\apj},
         year = 2015,
       volume = {815},
       number = {2},
        pages = {82},
          doi = {10.1088/0004-637X/815/2/82},
       adsurl = {https://ui.adsabs.harvard.edu/abs/2015ApJ...815...82L},
}

@ARTICLE{Mumpower2020,
       author = {{Mumpower}, M.~R. and {Jaffke}, P. and {Verriere}, M. and {Randrup}, J.},
        title = "{Primary fission fragment mass yields across the chart of nuclides}",
      journal = {\prc},
         year = 2020,
       volume = {101},
       number = {5},
        pages = {054607},
          doi = {10.1103/PhysRevC.101.054607},
       adsurl = {https://ui.adsabs.harvard.edu/abs/2020PhRvC.101e4607M},
}

@ARTICLE{Mumpower2022,
       author = {{Mumpower}, M.~R. and {Kawano}, T. and {Sprouse}, T.~M.},
        title = "{$\beta^{-}$-delayed fission in the coupled quasiparticle
                  random-phase approximation plus Hauser--Feshbach approach}",
      journal = {\prc},
         year = 2022,
       volume = {106},
       number = {6},
        pages = {065805},
          doi = {10.1103/PhysRevC.106.065805},
       adsurl = {https://ui.adsabs.harvard.edu/abs/2022PhRvC.106f5805M},
}

@ARTICLE{Oda1994,
       author = {{Oda}, T. and {Hino}, M. and {Muto}, K. and {Takahara}, M. and {Sato}, K.},
        title = "{Rate Tables for the Weak Processes of sd-Shell Nuclei in
                  Stellar Matter}",
      journal = {Atomic Data and Nuclear Data Tables},
         year = 1994,
       volume = {56},
       number = {2},
        pages = {231-403},
          doi = {10.1006/adnd.1994.1007},
       adsurl = {https://ui.adsabs.harvard.edu/abs/1994ADNDT..56..231O},
}

@ARTICLE{Pruet2003,
       author = {{Pruet}, Jason and {Fuller}, George M.},
        title = "{Estimates of Stellar Weak Interaction Rates for Nuclei in the
                  Mass Range A = 65--80}",
      journal = {\apjs},
         year = 2003,
       volume = {149},
       number = {1},
        pages = {189-203},
          doi = {10.1086/376753},
       adsurl = {https://ui.adsabs.harvard.edu/abs/2003ApJS..149..189P},
}

@ARTICLE{Suzuki2016,
       author = {{Suzuki}, Toshio and {Toki}, Hiroshi and {Nomoto}, Ken'ichi},
        title = "{Electron-capture and $\beta$-decay Rates for sd-Shell Nuclei
                  in Stellar Environments Relevant to High-density O-Ne-Mg Cores}",
      journal = {\apj},
         year = 2016,
       volume = {817},
       number = {2},
        pages = {163},
          doi = {10.3847/0004-637X/817/2/163},
       adsurl = {https://ui.adsabs.harvard.edu/abs/2016ApJ...817..163S},
}

@ARTICLE{Timmes1999,
       author = {{Timmes}, F.~X. and {Arnett}, Dave},
        title = "{The Accuracy, Consistency, and Speed of Five Equations of
                  State for Stellar Hydrodynamics}",
      journal = {\apjs},
         year = 1999,
       volume = {125},
       number = {1},
        pages = {277-294},
          doi = {10.1086/313271},
       adsurl = {https://ui.adsabs.harvard.edu/abs/1999ApJS..125..277T},
}

@ARTICLE{Setzer23,
       author = {{Setzer}, Christian N. and {Peiris}, Hiranya V. and
                 {Korobkin}, Oleg and {Rosswog}, Stephan},
        title = "{Modelling populations of kilonovae}",
      journal = {\mnras},
         year = 2023,
       volume = {520},
       number = {2},
        pages = {2829-2842},
          doi = {10.1093/mnras/stad257},
       adsurl = {https://ui.adsabs.harvard.edu/abs/2023MNRAS.520.2829S},
}

@misc{Biswas26,
      author        = {{Biswas}, Bhaskar and {Rosswog}, Stephan and
                       {Diener}, Peter and {Schnabel}, Lukas},
      title         = "{Binary neutron star mergers with {SPHINCS\_BSSN}:
                        temperature-dependent equations of state and
                        damping of constraint violations}",
      year          = {2026},
      eprint        = {2601.01402},
      archivePrefix = {arXiv},
      primaryClass  = {astro-ph.HE},
      url           = {https://arxiv.org/abs/2601.01402},
}

@ARTICLE{Wrean94,
       author = {{Wrean}, P.~R. and {Brune}, C.~R. and {Kavanagh}, R.~W.},
        title = "{Total cross sections and thermonuclear reaction rates for
                  $^{9}$Be($\alpha$,n)$^{12}$C}",
      journal = {\prc},
         year = 1994,
       volume = {49},
       number = {2},
        pages = {1205-1213},
          doi = {10.1103/PhysRevC.49.1205},
       adsurl = {https://ui.adsabs.harvard.edu/abs/1994PhRvC..49.1205W},
}

@ARTICLE{Shenhar24,
       author = {{Shenhar}, Ben and {Guttman}, Or and {Waxman}, Eli},
        title = "{An analytic description of electron thermalization in kilonovae ejecta}",
      journal = {\mnras},
         year = 2024,
        month = may,
       volume = {531},
       number = {1},
        pages = {592-601},
          doi = {10.1093/mnras/stae1218},
       adsurl = {https://ui.adsabs.harvard.edu/abs/2024MNRAS.531..592S},
}

@book{Longair11,
  author    = {Longair, Malcolm S.},
  title     = {High Energy Astrophysics},
  edition   = {3},
  year      = {2011},
  publisher = {Cambridge University Press},
  address   = {Cambridge},
  isbn      = {978-0-521-75618-1}
}

@article{Solodov08,
    author = {Solodov, A. A. and Betti, R.},
    title = {Stopping power and range of energetic electrons in dense plasmas of fast-ignition fusion targets},
    journal = {Physics of Plasmas},
    volume = {15},
    number = {4},
    pages = {042707},
    year = {2008},
    month = {04},
    issn = {1070-664X},
    doi = {10.1063/1.2903890},
    url = {https://doi.org/10.1063/1.2903890},
    eprint = {https://pubs.aip.org/aip/pop/article-pdf/doi/10.1063/1.2903890/13598330/042707_1_online.pdf},
}

@book{Segre77,
  author    = {Segre, E.},
  year      = {1977},
  title     = {Nuclei and Particles. An Introduction to Nuclear and Subnuclear Physics},
  edition   = {2nd},
  publisher = {W. A. Benjamin} 
}

@Preamble{ {\hyphenation{Post-Script Sprin-ger}}
}

@ARTICLE{aguilera24,
       author = {{Aguilera-Miret}, Ricard and {Palenzuela}, Carlos and {Carrasco}, Federico and {Rosswog}, Stephan and {Vigan{\`o}}, Daniele},
        title = "{Delayed jet launching in binary neutron star mergers with realistic initial magnetic fields}",
      journal = {\prd},
         year = 2024,
        month = oct,
       volume = {110},
       number = {8},
          eid = {083014},
        pages = {083014},
          doi = {10.1103/PhysRevD.110.083014},
archivePrefix = {arXiv},
       eprint = {2407.20335},
 primaryClass = {astro-ph.HE},
       adsurl = {https://ui.adsabs.harvard.edu/abs/2024PhRvD.110h3014A}
}

@ARTICLE{aguilera25,
       author = {{Aguilera-Miret}, Ricard and {Christian}, Jan-Erik and {Rosswog}, Stephan and {Palenzuela}, Carlos},
        title = "{Robustness of Magnetic Field Amplification in Neutron Star Mergers}",
      journal = {arXiv e-prints},
         year = 2025,
        month = apr,
          eid = {arXiv:2504.10604},
        pages = {arXiv:2504.10604},
          doi = {10.48550/arXiv.2504.10604},
archivePrefix = {arXiv},
       eprint = {2504.10604},
 primaryClass = {astro-ph.HE},
       adsurl = {https://ui.adsabs.harvard.edu/abs/2025arXiv250410604A}
}

@BOOK{alcubierre08,
   author = {{Alcubierre}, M.},
    title = "{Introduction to 3+1 Numerical Relativity}",
publisher = {Oxford University Press},
     year = 2008
}

@ARTICLE{baiotti17,
       author = {{Baiotti}, Luca and {Rezzolla}, Luciano},
        title = "{Binary neutron star mergers: a review of Einstein{\textquoteright}s richest laboratory}",
      journal = {Reports on Progress in Physics},
         year = "2017",
        month = "Sep",
       volume = {80},
       number = {9},
          eid = {096901},
        pages = {096901},
          doi = {10.1088/1361-6633/aa67bb},
archivePrefix = {arXiv},
       eprint = {1607.03540},
 primaryClass = {gr-qc},
       adsurl = {https://ui.adsabs.harvard.edu/abs/2017RPPh...80i6901B}
}

@ARTICLE{barnes16a,
   author = {{Barnes}, J. and {Kasen}, D. and {Wu}, M.-R. and {Martinez-Pinedo}, G.
	},
    title = "{Radioactivity and Thermalization in the Ejecta of Compact Object Mergers and Their Impact on Kilonova Light Curves}",
  journal = {ApJ},
archivePrefix = "arXiv",
   eprint = {1605.07218},
 primaryClass = "astro-ph.HE",
     year = 2016,
    month = oct,
   volume = 829,
      eid = {110},
    pages = {110},
      doi = {10.3847/0004-637X/829/2/110},
   adsurl = {http://adsabs.harvard.edu/abs/2016ApJ...829..110B}
}

@BOOK{baumgarte10,
   author = {{Baumgarte}, T.~W. and {Shapiro}, S.~L.},
    title = "{Numerical Relativity: Solving Einstein's Equations on the Computer}",
booktitle = {Numerical Relativity: Solving Einstein's Equations on the Computer},
     year = 2010,
     address= {Cambridge},
     publisher = {Cambridge University Press, ISBN: 9780521514071},
   adsurl = {http://adsabs.harvard.edu/abs/2010nure.book.....B}
}

@ARTICLE{bauswein13a,
   author = {{Bauswein}, A. and {Goriely}, S. and {Janka}, H.-T.},
    title = "{Systematics of Dynamical Mass Ejection, Nucleosynthesis, and Radioactively Powered Electromagnetic Signals from Neutron-star Mergers}",
  journal = {ApJ},
 primaryClass = "astro-ph.SR",
     year = 2013,
    month = aug,
   volume = 773,
      eid = {78},
    pages = {78},
      doi = {10.1088/0004-637X/773/1/78},
   adsurl = {http://adsabs.harvard.edu/abs/2013ApJ...773...78B}
}

@BOOK{baym04,
   author = {G. Baym and C. Pethick},
    title = "{Landau Fermi-Liquid Theory: Concepts and applications}",
publisher = {Wiley},
     year = 2004
}

@ARTICLE{bernuzzi25,
       author = {{Bernuzzi}, Sebastiano and {Magistrelli}, Fabio and {Jacobi}, Maximilian and {Logoteta}, Domenico and {Perego}, Albino and {Radice}, David},
        title = "{Long-lived neutron star remnants from asymmetric binary neutron star mergers: element formation, kilonova signals and gravitational waves}",
      journal = {MNRAS},
         year = 2025,
        month = sep,
       volume = {542},
       number = {1},
        pages = {256-271},
          doi = {10.1093/mnras/staf1147},
archivePrefix = {arXiv},
       eprint = {2409.18185},
 primaryClass = {astro-ph.HE},
       adsurl = {https://ui.adsabs.harvard.edu/abs/2025MNRAS.542..256B}
}

@ARTICLE{biswas22b,
       author = {{Biswas}, Bhaskar},
        title = "{Bayesian Model Selection of Neutron Star Equations of State Using Multi-messenger Observations}",
      journal = {ApJ},
         year = 2022,
        month = feb,
       volume = {926},
       number = {1},
          eid = {75},
        pages = {75},
          doi = {10.3847/1538-4357/ac447b},
archivePrefix = {arXiv},
       eprint = {2106.02644},
 primaryClass = {astro-ph.HE},
       adsurl = {https://ui.adsabs.harvard.edu/abs/2022ApJ...926...75B}
}

@ARTICLE{constantinou15,
       author = {{Constantinou}, Constantinos and {Muccioli}, Brian and {Prakash}, Madappa and {Lattimer}, James M.},
        title = "{Thermal properties of hot and dense matter with finite range interactions}",
      journal = {Phys. Rev. C},
         year = 2015,
        month = aug,
       volume = {92},
       number = {2},
          eid = {025801},
        pages = {025801},
          doi = {10.1103/PhysRevC.92.025801},
archivePrefix = {arXiv},
       eprint = {1504.03982},
 primaryClass = {astro-ph.SR},
       adsurl = {https://ui.adsabs.harvard.edu/abs/2015PhRvC..92b5801C}
}

@ARTICLE{cook26,
       author = {{Cook}, William and {Guti{\'e}rrez}, Eduardo M. and {Bernuzzi}, Sebastiano and {Radice}, David and {Daszuta}, Boris and {Fields}, Jacob and {Hammond}, Peter and {Bandyopadhyay}, Harshraj and {Jacobi}, Maximilian},
        title = "{Magnetic Field Configurations in Binary Neutron Star Mergers II: Inspiral, Merger and Ejecta}",
      journal = {arXiv e-prints},
         year = 2025,
        month = aug,
          eid = {arXiv:2508.19342},
        pages = {arXiv:2508.19342},
          doi = {10.48550/arXiv.2508.19342},
archivePrefix = {arXiv},
       eprint = {2508.19342},
 primaryClass = {astro-ph.HE},
       adsurl = {https://ui.adsabs.harvard.edu/abs/2025arXiv250819342C}
}

@ARTICLE{cowan21,
       author = {{Cowan}, John J. and {Sneden}, Christopher and {Lawler}, James E. and {Aprahamian}, Ani and {Wiescher}, Michael and {Langanke}, Karlheinz and {Martinez-Pinedo}, Gabriel and {Thielemann}, Friedrich-Karl},
        title = "{Origin of the heaviest elements: The rapid neutron-capture process}",
      journal = {Reviews of Modern Physics},
         year = 2021,
        month = jan,
       volume = {93},
       number = {1},
          eid = {015002},
        pages = {015002},
          doi = {10.1103/RevModPhys.93.015002},
archivePrefix = {arXiv},
       eprint = {1901.01410},
 primaryClass = {astro-ph.HE},
       adsurl = {https://ui.adsabs.harvard.edu/abs/2021RvMP...93a5002C}
}

@ARTICLE{dean21,
       author = {{Dean}, Coleman and {Fern{\'a}ndez}, Rodrigo and {Metzger}, Brian D.},
        title = "{Resolving the Fastest Ejecta from Binary Neutron Star Mergers: Implications for Electromagnetic Counterparts}",
      journal = {ApJ},
         year = 2021,
        month = nov,
       volume = {921},
       number = {2},
          eid = {161},
        pages = {161},
          doi = {10.3847/1538-4357/ac1f20},
archivePrefix = {arXiv},
       eprint = {2108.08311},
 primaryClass = {astro-ph.HE},
       adsurl = {https://ui.adsabs.harvard.edu/abs/2021ApJ...921..161D}
}

@ARTICLE{diener22a,
       author = {{Diener}, Peter and {Rosswog}, Stephan and {Torsello}, Francesco},
        title = "{Simulating neutron star mergers with the Lagrangian Numerical Relativity code SPHINCS\_BSSN}",
      journal = {European Physical Journal A},
         year = 2022,
        month = apr,
       volume = {58},
       number = {4},
          eid = {74},
        pages = {74},
          doi = {10.1140/epja/s10050-022-00725-7},
archivePrefix = {arXiv},
       eprint = {2203.06478},
 primaryClass = {astro-ph.HE},
       adsurl = {https://ui.adsabs.harvard.edu/abs/2022EPJA...58...74D}
}

@ARTICLE{engvik96,
       author = {{Engvik}, L. and {Osnes}, E. and {Hjorth-Jensen}, M. and {Bao}, G. and {Ostgaard}, E.},
        title = "{Asymmetric Nuclear Matter and Neutron Star Properties}",
      journal = {ApJ},
         year = 1996,
        month = oct,
       volume = {469},
        pages = {794},
          doi = {10.1086/177827},
archivePrefix = {arXiv},
       eprint = {nucl-th/9509016},
 primaryClass = {nucl-th},
       adsurl = {https://ui.adsabs.harvard.edu/abs/1996ApJ...469..794E}
}

@ARTICLE{farouqi10,
   author = {{Farouqi}, K. and {Kratz}, K.-L. and {Pfeiffer}, B. and {Rauscher}, T. and 
	{Thielemann}, F.-K. and {Truran}, J.~W.},
    title = "{Charged-particle and Neutron-capture Processes in the High-entropy Wind of Core-collapse Supernovae}",
  journal = {ApJ},
archivePrefix = "arXiv",
   eprint = {1002.2346},
 primaryClass = "astro-ph.SR",
     year = 2010,
    month = apr,
   volume = 712,
    pages = {1359-1377},
      doi = {10.1088/0004-637X/712/2/1359},
   adsurl = {http://adsabs.harvard.edu/abs/2010ApJ...712.1359F}
}

@ARTICLE{farouqi22,
       author = {{Farouqi}, K. and {Thielemann}, F. -K. and {Rosswog}, S. and {Kratz}, K. -L.},
        title = "{Correlations of r-process elements in very metal-poor stars as clues to their nucleosynthesis sites}",
      journal = {\aap},
         year = 2022,
        month = jul,
       volume = {663},
          eid = {A70},
        pages = {A70},
          doi = {10.1051/0004-6361/202141038},
archivePrefix = {arXiv},
       eprint = {2107.03486},
 primaryClass = {astro-ph.SR},
       adsurl = {https://ui.adsabs.harvard.edu/abs/2022A&A...663A..70F}
}

@ARTICLE{foucart21b,
       author = {{Foucart}, Francois and {M{\"o}sta}, Philipp and {Ramirez}, Teresita and {Wright}, Alex James and {Darbha}, Siva and {Kasen}, Daniel},
        title = "{Estimating outflow masses and velocities in merger simulations: Impact of r -process heating and neutrino cooling}",
      journal = {\prd},
         year = 2021,
        month = dec,
       volume = {104},
       number = {12},
          eid = {123010},
        pages = {123010},
          doi = {10.1103/PhysRevD.104.123010},
archivePrefix = {arXiv},
       eprint = {2109.00565},
 primaryClass = {astro-ph.HE},
       adsurl = {https://ui.adsabs.harvard.edu/abs/2021PhRvD.104l3010F}
}

@ARTICLE{foucart23,
       author = {{Foucart}, Francois},
        title = "{Neutrino transport in general relativistic neutron star merger simulations}",
      journal = {Living Reviews in Computational Astrophysics},
         year = 2023,
        month = dec,
       volume = {9},
       number = {1},
          eid = {1},
        pages = {1},
          doi = {10.1007/s41115-023-00016-y},
archivePrefix = {arXiv},
       eprint = {2209.02538},
 primaryClass = {astro-ph.HE},
       adsurl = {https://ui.adsabs.harvard.edu/abs/2023LRCA....9....1F}
}

@ARTICLE{freiburghaus99a,
 author = {C. Freiburghaus and J.F. Rembges and T. Rauscher and E. Kolbe
and F.-K. Thielemann and K.-L. Kratz and J.J. Cowan},
 title = {The Astrophysical r-Process: A Comparison of Calculations following Adiabatic Expansion with Classical Calculations Based on Neutron Densities and Temperatures},
 journal = {ApJ},
 year = {1999},
 volume = {516},
 pages = {381}}

@ARTICLE{fuller95,
 author = {G. Fuller and B. S. Meyer},
 title = {},
 journal = {ApJ},
 year = {1995},
 volume = {453},
 pages = {792}}

@ARTICLE{gottlieb23,
       author = {{Gottlieb}, Ore and {Metzger}, Brian D. and {Quataert}, Eliot and {Issa}, Danat and {Martineau}, Tia and {Foucart}, Francois and {Duez}, Matthew D. and {Kidder}, Lawrence E. and {Pfeiffer}, Harald P. and {Scheel}, Mark A.},
        title = "{A Unified Picture of Short and Long Gamma-Ray Bursts from Compact Binary Mergers}",
      journal = {\apjl},
         year = 2023,
        month = dec,
       volume = {958},
       number = {2},
          eid = {L33},
        pages = {L33},
          doi = {10.3847/2041-8213/ad096e},
archivePrefix = {arXiv},
       eprint = {2309.00038},
 primaryClass = {astro-ph.HE},
       adsurl = {https://ui.adsabs.harvard.edu/abs/2023ApJ...958L..33G}
}

@ARTICLE{hajela22,
       author = {{Hajela}, A. and {Margutti}, R. and {Bright}, J.~S. and {Alexander}, K.~D. and {Metzger}, B.~D. and {Nedora}, V. and {Kathirgamaraju}, A. and {Margalit}, B. and {Radice}, D. and {Guidorzi}, C. and {Berger}, E. and {MacFadyen}, A. and {Giannios}, D. and {Chornock}, R. and {Heywood}, I. and {Sironi}, L. and {Gottlieb}, O. and {Coppejans}, D. and {Laskar}, T. and {Cendes}, Y. and {Duran}, R. Barniol and {Eftekhari}, T. and {Fong}, W. and {McDowell}, A. and {Nicholl}, M. and {Xie}, X. and {Zrake}, J. and {Bernuzzi}, S. and {Broekgaarden}, F.~S. and {Kilpatrick}, C.~D. and {Terreran}, G. and {Villar}, V.~A. and {Blanchard}, P.~K. and {Gomez}, S. and {Hosseinzadeh}, G. and {Matthews}, D.~J. and {Rastinejad}, J.~C.},
        title = "{Evidence for X-Ray Emission in Excess to the Jet-afterglow Decay 3.5 yr after the Binary Neutron Star Merger GW 170817: A New Emission Component}",
      journal = {ApJL},
         year = 2022,
        month = mar,
       volume = {927},
       number = {1},
          eid = {L17},
        pages = {L17},
          doi = {10.3847/2041-8213/ac504a},
archivePrefix = {arXiv},
       eprint = {2104.02070},
 primaryClass = {astro-ph.HE},
       adsurl = {https://ui.adsabs.harvard.edu/abs/2022ApJ...927L..17H}
}

@ARTICLE{hoffman97,
 author = {R. D. Hoffman and S. E. Woosley and Y.-Z. Qian},
 title = {{N}ucleosynthesis in {N}eutrino-{D}riven {W}inds. II. {I}mplications
		  for heavy {E}lement {S}ynthesis},
 journal = {ApJ},
 year = {1997},
 volume = {482},
 pages = {951}}

@ARTICLE{hotokezaka18a,
       author = {{Hotokezaka}, Kenta and {Kiuchi}, Kenta and {Shibata}, Masaru and {Nakar}, Ehud and {Piran}, Tsvi},
        title = "{Synchrotron Radiation from the Fast Tail of Dynamical Ejecta of Neutron Star Mergers}",
      journal = {ApJ},
         year = 2018,
        month = nov,
       volume = {867},
       number = {2},
          eid = {95},
        pages = {95},
          doi = {10.3847/1538-4357/aadf92},
archivePrefix = {arXiv},
       eprint = {1803.00599},
 primaryClass = {astro-ph.HE},
       adsurl = {https://ui.adsabs.harvard.edu/abs/2018ApJ...867...95H}
}

@ARTICLE{hotokezaka13,
   author = {{Hotokezaka}, K. and {Kiuchi}, K. and {Kyutoku}, K. and {Okawa}, H. and 
	{Sekiguchi}, Y.-i. and {Shibata}, M. and {Taniguchi}, K.},
    title = "{Mass ejection from the merger of binary neutron stars}",
  journal = {\prd},
archivePrefix = "arXiv",
   eprint = {1212.0905},
 primaryClass = "astro-ph.HE",
     year = 2013,
    month = jan,
   volume = 87,
   number = 2,
      eid = {024001},
    pages = {024001},
      doi = {10.1103/PhysRevD.87.024001},
   adsurl = {http://adsabs.harvard.edu/abs/2013PhRvD..87b4001H}
}

@ARTICLE{hotokezaka17a,
   author = {{Hotokezaka}, K. and {Sari}, R. and {Piran}, T.},
    title = "{Analytic heating rate of neutron star merger ejecta derived from Fermi's theory of beta decay}",
  journal = {MNRAS},
archivePrefix = "arXiv",
   eprint = {1701.02785},
 primaryClass = "astro-ph.HE",
     year = 2017,
    month = jun,
   volume = 468,
    pages = {91-96},
      doi = {10.1093/mnras/stx411},
   adsurl = {http://adsabs.harvard.edu/abs/2017MNRAS.468...91H}
}

@ARTICLE{kasen13a,
   author = {{Kasen}, D. and {Badnell}, N.~R. and {Barnes}, J.},
    title = "{Opacities and Spectra of the r-process Ejecta from Neutron Star Mergers}",
  journal = {ApJ},
archivePrefix = "arXiv",
   eprint = {1303.5788},
 primaryClass = "astro-ph.HE",
     year = 2013,
    month = sep,
   volume = 774,
      eid = {25},
    pages = {25},
      doi = {10.1088/0004-637X/774/1/25},
   adsurl = {http://adsabs.harvard.edu/abs/2013ApJ...774...25K}
}

@ARTICLE{korobkin12a,
   author = {{Korobkin}, O. and {Rosswog}, S. and {Arcones}, A. and {Winteler}, C.},
    title = "{On the astrophysical robustness of the neutron star merger r-process}",
  journal = {MNRAS},
 primaryClass = "astro-ph.SR",
     year = 2012,
    month = nov,
   volume = 426,
    pages = {1940-1949}
}

@ARTICLE{kumar15,
   author = {{Kumar}, P. and {Zhang}, B.},
    title = "{The physics of gamma-ray bursts \& relativistic jets}",
  journal = {Phys. Rep.},
archivePrefix = "arXiv",
   eprint = {1410.0679},
 primaryClass = "astro-ph.HE",
     year = 2015,
    month = feb,
   volume = 561,
    pages = {1-109},
      doi = {10.1016/j.physrep.2014.09.008},
   adsurl = {http://adsabs.harvard.edu/abs/2015PhR...561....1K}
}

@ARTICLE{langanke01,
   author = {{Langanke}, K. and {Martinez-Pinedo}, G.},
    title = "{Rate Tables for the Weak Processes of pf-SHELL Nuclei in Stellar Environments}",
  journal = {Atomic Data and Nuclear Data Tables},
     year = 2001,
    month = sep,
   volume = 79,
    pages = {1-46},
      doi = {10.1006/adnd.2001.0865},
   adsurl = {http://adsabs.harvard.edu/abs/2001ADNDT..79....1L}
}

@ARTICLE{lee07,
   author = {{Lee}, W.~H. and {Ramirez-Ruiz}, E.},
    title = "{The progenitors of short gamma-ray bursts}",
  journal = {New Journal of Physics},
   eprint = {arXiv:astro-ph/0701874},
     year = 2007,
    month = jan,
   volume = 9,
    pages = {17},
      doi = {10.1088/1367-2630/9/1/017},
   adsurl = {http://adsabs.harvard.edu/abs/2007NJPh....9...17L}
}

@ARTICLE{matur24,
       author = {{Matur}, Rahime and {Hawke}, Ian and {Andersson}, Nils},
        title = "{Signatures of low-mass black hole-neutron star mergers}",
      journal = {\mnras},
         year = 2024,
        month = nov,
       volume = {534},
       number = {3},
        pages = {2894-2903},
          doi = {10.1093/mnras/stae2238},
archivePrefix = {arXiv},
       eprint = {2407.18045},
 primaryClass = {astro-ph.HE},
       adsurl = {https://ui.adsabs.harvard.edu/abs/2024MNRAS.534.2894M}
}

@ARTICLE{meszaros06,
   author = {{Meszaros}, P.},
    title = "{Gamma-ray bursts.}",
  journal = {Reports of Progress in Physics},
   eprint = {astro-ph/0605208},
     year = 2006,
   volume = 69,
    pages = {2259-2322},
   adsurl = {http://adsabs.harvard.edu/cgi-bin/nph-bib_query?bibcode=2006RPPh...69.2259M&db_key=AST}
}

@ARTICLE{metzger10b,
   author = {{Metzger}, B.~D. and {Martinez-Pinedo}, G. and {Darbha}, S. and 
	{Quataert}, E. and {Arcones}, A. and {Kasen}, D. and {Thomas}, R. and 
	{Nugent}, P. and {Panov}, I.~V. and {Zinner}, N.~T.},
    title = "{Electromagnetic counterparts of compact object mergers powered by the radioactive decay of r-process nuclei}",
  journal = {MNRAS},
archivePrefix = "arXiv",
   eprint = {1001.5029},
 primaryClass = "astro-ph.HE",
     year = 2010,
    month = aug,
   volume = 406,
    pages = {2650-2662},
      doi = {10.1111/j.1365-2966.2010.16864.x},
   adsurl = {http://adsabs.harvard.edu/abs/2010MNRAS.406.2650M}
}

@ARTICLE{metzger15a,
   author = {{Metzger}, B.~D. and {Bauswein}, A. and {Goriely}, S. and {Kasen}, D.},
    title = "{Neutron-powered precursors of kilonovae}",
  journal = {MNRAS},
archivePrefix = "arXiv",
   eprint = {1409.0544},
 primaryClass = "astro-ph.HE",
     year = 2015,
    month = jan,
   volume = 446,
    pages = {1115-1120},
      doi = {10.1093/mnras/stu2225},
   adsurl = {http://adsabs.harvard.edu/abs/2015MNRAS.446.1115M}
}

@ARTICLE{metzger20,
       author = {{Metzger}, Brian D.},
        title = "{Kilonovae}",
      journal = {Living Reviews in Relativity},
         year = 2020,
        month = dec,
       volume = {23},
       number = {1},
          eid = {1},
        pages = {1},
          doi = {10.1007/s41114-019-0024-0},
       adsurl = {https://ui.adsabs.harvard.edu/abs/2020LRR....23....1M}
}

@ARTICLE{moeller95,
 author = {P. M\"oller and J. R. Nix and W. D. Myers and W. J. Swiatecki},
 title = {},
 journal = {At. Data Nucl. Data Tables},
 year = {1995},
 volume = {59},
 pages = {185}}

@ARTICLE{mooley17,
   author = {{Mooley}, K.~P. and {Nakar}, E. and {Hotokezaka}, K. and {Hallinan}, G. and 
	{Corsi}, A. and {Frail}, D.~A. and {Horesh}, A. and {Murphy}, T. and 
	{Lenc}, E. and {Kaplan}, D.~L. and {de}, K. and {Dobie}, D. and 
	{Chandra}, P. and {Deller}, A. and {Gottlieb}, O. and {Kasliwal}, M.~M. and 
	{Kulkarni}, S.~R. and {Myers}, S.~T. and {Nissanke}, S. and 
	{Piran}, T. and {Lynch}, C. and {Bhalerao}, V. and {Bourke}, S. and 
	{Bannister}, K.~W. and {Singer}, L.~P.},
    title = "{A mildly relativistic wide-angle outflow in the neutron-star merger event GW170817}",
  journal = {Nature},
archivePrefix = "arXiv",
   eprint = {1711.11573},
 primaryClass = "astro-ph.HE",
     year = 2018,
    month = feb,
   volume = 554,
    pages = {207-210},
      doi = {10.1038/nature25452},
   adsurl = {http://adsabs.harvard.edu/abs/2018Natur.554..207M}
}

@ARTICLE{nakar07,
   author = {{Nakar}, E.},
    title = "{Short-hard gamma-ray bursts}",
  journal = {Phys. Rep.},
   eprint = {arXiv:astro-ph/0701748},
     year = 2007,
    month = apr,
   volume = 442,
    pages = {166-236},
      doi = {10.1016/j.physrep.2007.02.005},
   adsurl = {http://adsabs.harvard.edu/abs/2007PhR...442..166N}
}

@ARTICLE{nakar11a,
   author = {{Nakar}, E. and {Piran}, T.},
    title = "{Detectable radio flares following gravitational waves from mergers of binary neutron stars}",
  journal = {Nature},
     year = 2011,
    month = oct,
   volume = 478,
    pages = {82-84},
      doi = {10.1038/nature10365},
   adsurl = {http://adsabs.harvard.edu/abs/2011Natur.478...82N}
}

@ARTICLE{nedora21,
       author = {{Nedora}, Vsevolod and {Radice}, David and {Bernuzzi}, Sebastiano and {Perego}, Albino and {Daszuta}, Boris and {Endrizzi}, Andrea and {Prakash}, Aviral and {Schianchi}, Federico},
        title = "{Dynamical ejecta synchrotron emission as a possible contributor to the changing behaviour of GRB170817A afterglow}",
      journal = {MNRAS},
         year = 2021,
        month = oct,
       volume = {506},
       number = {4},
        pages = {5908-5915},
          doi = {10.1093/mnras/stab2004},
archivePrefix = {arXiv},
       eprint = {2104.04537},
 primaryClass = {astro-ph.HE},
       adsurl = {https://ui.adsabs.harvard.edu/abs/2021MNRAS.506.5908N}
}

@ARTICLE{neuweiler23,
       author = {{Neuweiler}, Anna and {Dietrich}, Tim and {Bulla}, Mattia and {Chaurasia}, Swami Vivekanandji and {Rosswog}, Stephan and {Ujevic}, Maximiliano},
        title = "{Long-term simulations of dynamical ejecta: Homologous expansion and kilonova properties}",
      journal = {Phys. Rev. D},
         year = 2023,
        month = jan,
       volume = {107},
       number = {2},
          eid = {023016},
        pages = {023016},
          doi = {10.1103/PhysRevD.107.023016},
archivePrefix = {arXiv},
       eprint = {2208.13460},
 primaryClass = {astro-ph.HE},
       adsurl = {https://ui.adsabs.harvard.edu/abs/2023PhRvD.107b3016N}
}

@ARTICLE{neuweiler26,
       author = {{Neuweiler}, Anna and {Gieg}, Henrique and {Rose}, Henrik and {Koehn}, Hauke and {Markin}, Ivan and {Schianchi}, Federico and {Brodie}, Liam and {Haber}, Alexander and {Nedora}, Vsevolod and {Bulla}, Mattia and {Dietrich}, Tim},
        title = "{General-relativistic radiation magnetohydrodynamics simulations of binary neutron star mergers: The influence of spin on the multimessenger picture}",
      journal = {\prd},
         year = 2026,
        month = feb,
       volume = {113},
       number = {4},
          eid = {043038},
        pages = {043038},
          doi = {10.1103/mxlf-8sbm},
archivePrefix = {arXiv},
       eprint = {2510.14850},
 primaryClass = {astro-ph.HE},
       adsurl = {https://ui.adsabs.harvard.edu/abs/2026PhRvD.113d3038N}
}

@ARTICLE{panov10,
   author = {{Panov}, I.~V. and {Korneev}, I.~Y. and {Rauscher}, T. and {Martinez-Pinedo}, G. and 
	{Keli{\'c}-Heil}, A. and {Zinner}, N.~T. and {Thielemann}, F.-K.
	},
    title = "{Neutron-induced astrophysical reaction rates for translead nuclei}",
  journal = {A \& A},
archivePrefix = "arXiv",
   eprint = {0911.2181},
 primaryClass = "astro-ph.SR",
     year = 2010,
    month = apr,
   volume = 513,
      eid = {A61},
    pages = {A61},
      doi = {10.1051/0004-6361/200911967},
   adsurl = {http://adsabs.harvard.edu/abs/2010A%26A...513A..61P}
}

@ARTICLE{perego19a,
       author = {{Perego}, Albino and {Bernuzzi}, Sebastiano and {Radice}, David},
        title = "{Thermodynamics conditions of matter in neutron star mergers}",
      journal = {European Physical Journal A},
         year = 2019,
        month = aug,
       volume = {55},
       number = {8},
          eid = {124},
          pages = {124},
          doi = {10.1140/epja/i2019-12810-7},
archivePrefix = {arXiv},
       eprint = {1903.07898},
 primaryClass = {gr-qc},
       adsurl = {https://ui.adsabs.harvard.edu/abs/2019EPJA...55..124P}
}

@ARTICLE{piran04,
   author = {{Piran}, T.},
    title = "{The physics of gamma-ray bursts}",
  journal = {Reviews of Modern Physics},
   eprint = {arXiv:astro-ph/0405503},
     year = 2004,
    month = oct,
   volume = 76,
    pages = {1143-1210},
      doi = {10.1103/RevModPhys.76.1143},
   adsurl = {http://adsabs.harvard.edu/abs/2004RvMP...76.1143P}
}

@ARTICLE{price06,
 author = {D.J. Price and S. Rosswog},
 title = {Producing ultra-strong magnetic fields in neutron star mergers},
 journal = {Science},
 year = {2006},
 volume = {312},
 pages = {719}}

@ARTICLE{radice18a,
       author = {{Radice}, David and {Perego}, Albino and {Hotokezaka}, Kenta and
         {Fromm}, Steven A. and {Bernuzzi}, Sebastiano and {Roberts}, Luke F.},
        title = "{Binary Neutron Star Mergers: Mass Ejection, Electromagnetic Counterparts, and Nucleosynthesis}",
      journal = {ApJ},
         year = "2018",
        month = "Dec",
       volume = {869},
       number = {2},
          eid = {130},
        pages = {130},
          doi = {10.3847/1538-4357/aaf054},
archivePrefix = {arXiv},
       eprint = {1809.11161},
 primaryClass = {astro-ph.HE},
       adsurl = {https://ui.adsabs.harvard.edu/abs/2018ApJ...869..130R}
}

@ARTICLE{raithel19,
       author = {{Raithel}, Carolyn A. and {{\"O}zel}, Feryal and {Psaltis}, Dimitrios},
        title = "{Finite-temperature Extension for Cold Neutron Star Equations of State}",
      journal = {ApJ},
         year = 2019,
        month = apr,
       volume = {875},
       number = {1},
          eid = {12},
        pages = {12},
          doi = {10.3847/1538-4357/ab08ea},
archivePrefix = {arXiv},
       eprint = {1902.10735},
 primaryClass = {astro-ph.HE},
       adsurl = {https://ui.adsabs.harvard.edu/abs/2019ApJ...875...12R}
}

@ARTICLE{raithel21a,
       author = {{Raithel}, Carolyn A. and {Paschalidis}, Vasileios and {{\"O}zel}, Feryal},
        title = "{Realistic finite-temperature effects in neutron star merger simulations}",
      journal = {Phys. Rev. D},
         year = 2021,
        month = sep,
       volume = {104},
       number = {6},
          eid = {063016},
        pages = {063016},
          doi = {10.1103/PhysRevD.104.063016},
archivePrefix = {arXiv},
       eprint = {2104.07226},
 primaryClass = {astro-ph.HE},
       adsurl = {https://ui.adsabs.harvard.edu/abs/2021PhRvD.104f3016R}
}

@ARTICLE{read09,
       author = {{Read}, Jocelyn S. and {Lackey}, Benjamin D. and {Owen}, Benjamin J. and {Friedman}, John L.},
        title = "{Constraints on a phenomenologically parametrized neutron-star equation of state}",
      journal = {Phys. Rev. D},
         year = 2009,
        month = jun,
       volume = {79},
       number = {12},
          eid = {124032},
        pages = {124032},
          doi = {10.1103/PhysRevD.79.124032},
archivePrefix = {arXiv},
       eprint = {0812.2163},
 primaryClass = {astro-ph},
       adsurl = {https://ui.adsabs.harvard.edu/abs/2009PhRvD..79l4032R}
}

@ARTICLE{reichert23,
       author = {{Reichert}, M. and {Winteler}, C. and {Korobkin}, O. and {Arcones}, A. and {Bliss}, J. and {Eichler}, M. and {Frischknecht}, U. and {Fr{\"o}hlich}, C. and {Hirschi}, R. and {Jacobi}, M. and {Kuske}, J. and {Mart{\'\i}nez-Pinedo}, G. and {Martin}, D. and {Mocelj}, D. and {Rauscher}, T. and {Thielemann}, F. -K.},
        title = "{The Nuclear Reaction Network WinNet}",
      journal = {\apjs},
         year = 2023,
        month = oct,
       volume = {268},
       number = {2},
          eid = {66},
        pages = {66},
          doi = {10.3847/1538-4365/acf033},
archivePrefix = {arXiv},
       eprint = {2305.07048},
 primaryClass = {astro-ph.IM},
       adsurl = {https://ui.adsabs.harvard.edu/abs/2023ApJS..268...66R}
}

@BOOK{rezzolla13a,
   author = {{Rezzolla}, L. and {Zanotti}, O.},
    title = "{Relativistic Hydrodynamics}",
booktitle = {Relativistic Hydrodynamics},
     year = 2013,
    month = "sep",
publisher = {Oxford University Press, 2013.~ISBN-10: 0198528906; ISBN-13: 978-0198528906},
   adsurl = {http://adsabs.harvard.edu/abs/2013rehy.book.....R}
}

@ARTICLE{roberts11,
   author = {{Roberts}, L.~F. and {Kasen}, D. and {Lee}, W.~H. and {Ramirez-Ruiz}, E.
	},
    title = "{Electromagnetic Transients Powered by Nuclear Decay in the Tidal Tails of Coalescing Compact Binaries}",
  journal = {ApJL},
 primaryClass = "astro-ph.HE",
     year = 2011,
    month = jul,
   volume = 736,
    pages = {L21}
}

@ARTICLE{rosswog03a,
   author = {{Rosswog}, S. and {Liebend{\"o}rfer}, M.},
    title = "{High-resolution calculations of merging neutron stars - II. Neutrino emission}",
  journal = {MNRAS},
     year = 2003,
    month = jul,
   volume = 342,
    pages = {673-689},
      doi = {10.1046/j.1365-8711.2003.06579.x},
   adsurl = {http://adsabs.harvard.edu/cgi-bin/nph-bib_query?bibcode=2003MNRAS.342..673R&db_key=AST}
}

@ARTICLE{rosswog05a,
   author = {{Rosswog}, S.},
    title = "{Mergers of Neutron Star-Black Hole Binaries with Small Mass Ratios: Nucleosynthesis, Gamma-Ray Bursts, and Electromagnetic Transients}",
  journal = {ApJ},
     year = 2005,
    month = dec,
   volume = 634,
    pages = {1202-1213},
      doi = {10.1086/497062},
   adsurl = {http://adsabs.harvard.edu/cgi-bin/nph-bib_query?bibcode=2005ApJ...634.1202R&db_key=AST}
}

@ARTICLE{rosswog14a,
   author = {{Rosswog}, S. and {Korobkin}, O. and {Arcones}, A. and {Thielemann}, F.-K. and 
	{Piran}, T.},
    title = "{The long-term evolution of neutron star merger remnants - I. The impact of r-process nucleosynthesis}",
  journal = {MNRAS},
archivePrefix = "arXiv",
   eprint = {1307.2939},
 primaryClass = "astro-ph.HE",
     year = 2014,
    month = mar,
   volume = 439,
    pages = {744-756},
      doi = {10.1093/mnras/stt2502},
   adsurl = {http://adsabs.harvard.edu/abs/2014MNRAS.439..744R}
}

@ARTICLE{rosswog18a,
   author = {{Rosswog}, S. and {Sollerman}, J. and {Feindt}, U. and {Goobar}, A. and 
	{Korobkin}, O. and {Wollaeger}, R. and {Fremling}, C. and {Kasliwal}, M.~M.
	},
    title = "{The first direct double neutron star merger detection: Implications for cosmic nucleosynthesis}",
  journal = {A\&A},
archivePrefix = "arXiv",
   eprint = {1710.05445},
 primaryClass = "astro-ph.HE",
     year = 2018,
    month = "jul",
   volume = 615,
      eid = {A132},
    pages = {A132},
      doi = {10.1051/0004-6361/201732117},
   adsurl = {http://adsabs.harvard.edu/abs/2018A%26A...615A.132R}
}

@ARTICLE{rosswog21a,
       author = {{Rosswog}, S. and {Diener}, P.},
        title = "{SPHINCS\_BSSN: a general relativistic smooth particle hydrodynamics code for dynamical spacetimes}",
      journal = {Classical and Quantum Gravity},
         year = 2021,
        month = jun,
       volume = {38},
       number = {11},
          eid = {115002},
        pages = {115002},
          doi = {10.1088/1361-6382/abee65},
archivePrefix = {arXiv},
       eprint = {2012.13954},
 primaryClass = {gr-qc},
       adsurl = {https://ui.adsabs.harvard.edu/abs/2021CQGra..38k5002R}
}

@ARTICLE{rosswog22b,
       author = {{Rosswog}, Stephan and {Diener}, Peter and {Torsello}, Francesco},
        title = "{Thinking Outside the Box: Numerical Relativity with Particles}",
      journal = {Symmetry},
         year = 2022,
        month = jun,
       volume = {14},
       number = {6},
        pages = {1280},
          doi = {10.3390/sym14061280},
archivePrefix = {arXiv},
       eprint = {2205.08130},
 primaryClass = {gr-qc},
       adsurl = {https://ui.adsabs.harvard.edu/abs/2022Symm...14.1280R}
}

@ARTICLE{rosswog23a,
       author = {{Rosswog}, Stephan and {Torsello}, Francesco and {Diener}, Peter},
        title = "{The Lagrangian Numerical Relativity code SPHINCS\_BSSN\_v1.0}",
      journal = {Front. Appl. Math. Stat.},
         year = 2023,
        month = jun,
       volume = {9},
          doi = {10.48550/arXiv.2306.06226},
archivePrefix = {arXiv},
       eprint = {2306.06226},
 primaryClass = {gr-qc},
       adsurl = {https://ui.adsabs.harvard.edu/abs/2023arXiv230606226R}
}

@ARTICLE{rosswog24b,
       author = {{Rosswog}, S. and {Diener}, P. and {Torsello}, F. and {Tauris}, T.~M. and {Sarin}, N.},
        title = "{Mergers of double NSs with one high-spin component: brighter kilonovae and fallback accretion, weaker gravitational waves}",
      journal = {MNRAS},
         year = 2024,
        month = may,
       volume = {530},
       number = {2},
        pages = {2336-2354},
          doi = {10.1093/mnras/stae454},
archivePrefix = {arXiv},
       eprint = {2310.15920},
 primaryClass = {astro-ph.HE},
       adsurl = {https://ui.adsabs.harvard.edu/abs/2024MNRAS.530.2336R}
}

@BOOK{rosswog25c,
  author = {Stephan Rosswog and Peter Diener},
  title     = {'SPHINCS\_BSSN: Numerical Relativity with Particles' in book 'New Frontiers in GRMHD Simulations'},
  series    = {Springer Series in Astrophysics and Cosmology},
  publisher = {Springer Nature Singapore},
  year      = {2025},
  url       = {https://link.springer.com/book/10.1007/978-981-97-8522-3},
  isbn      = {9819785219, 9789819785216}
}

@ARTICLE{rosswog25b,
       author = {{Rosswog}, Stephan and {Sarin}, Nikhil and {Nakar}, Ehud and {Diener}, Peter},
        title = "{Fast dynamic ejecta in neutron star mergers}",
      journal = {MNRAS},
         year = 2025,
        month = apr,
       volume = {538},
       number = {2},
        pages = {907-924},
          doi = {10.1093/mnras/staf324},
archivePrefix = {arXiv},
       eprint = {2411.18813},
 primaryClass = {astro-ph.HE},
       adsurl = {https://ui.adsabs.harvard.edu/abs/2025MNRAS.538..907R}
}

@ARTICLE{ruiz16,
   author = {{Ruiz}, M. and {Lang}, R.~N. and {Paschalidis}, V. and {Shapiro}, S.~L.
	},
    title = "{Binary Neutron Star Mergers: A Jet Engine for Short Gamma-Ray Bursts}",
  journal = {ApJL},
archivePrefix = "arXiv",
   eprint = {1604.02455},
 primaryClass = "astro-ph.HE",
     year = 2016,
    month = jun,
   volume = 824,
      eid = {L6},
    pages = {L6},
      doi = {10.3847/2041-8205/824/1/L6},
   adsurl = {http://adsabs.harvard.edu/abs/2016ApJ...824L...6R}
}

@ARTICLE{sadeh23,
       author = {{Sadeh}, Gilad and {Guttman}, Or and {Waxman}, Eli},
        title = "{Non-thermal emission from mildly relativistic dynamical ejecta of neutron star mergers}",
      journal = {MNRAS},
         year = 2023,
        month = jan,
       volume = {518},
       number = {2},
        pages = {2102-2112},
          doi = {10.1093/mnras/stac3260},
archivePrefix = {arXiv},
       eprint = {2207.05746},
 primaryClass = {astro-ph.HE},
       adsurl = {https://ui.adsabs.harvard.edu/abs/2023MNRAS.518.2102S}
}

@ARTICLE{sadeh24,
       author = {{Sadeh}, Gilad and {Linder}, Noya and {Waxman}, Eli},
        title = "{Non-thermal emission from mildly relativistic dynamical ejecta of neutron star mergers: spectrum and sky image}",
      journal = {arXiv e-prints},
         year = 2024,
        month = mar,
          eid = {arXiv:2403.07047},
        pages = {arXiv:2403.07047},
          doi = {10.48550/arXiv.2403.07047},
archivePrefix = {arXiv},
       eprint = {2403.07047},
 primaryClass = {astro-ph.HE},
       adsurl = {https://ui.adsabs.harvard.edu/abs/2024arXiv240307047S}
}

@article{sarin20,
    author = "Sarin, Nikhil and Lasky, Paul D.",
    title = "{The evolution of binary neutron star post-merger remnants: a review}",
    eprint = "2012.08172",
    archivePrefix = "arXiv",
    primaryClass = "astro-ph.HE",
    doi = "10.1007/s10714-021-02831-1",
    journal = "Gen. Rel. Grav.",
    volume = "53",
    number = "6",
    pages = "59",
    year = "2021"
}

@ARTICLE{sekiguchi11,
   author = {{Sekiguchi}, Y. and {Kiuchi}, K. and {Kyutoku}, K. and {Shibata}, M.
	},
    title = "{Gravitational Waves and Neutrino Emission from the Merger of Binary Neutron Stars}",
  journal = {Physical Review Letters},
archivePrefix = "arXiv",
   eprint = {1105.2125},
 primaryClass = "gr-qc",
     year = 2011,
    month = jul,
   volume = 107,
   number = 5,
      eid = {051102},
    pages = {051102},
      doi = {10.1103/PhysRevLett.107.051102},
   adsurl = {http://adsabs.harvard.edu/abs/2011PhRvL.107e1102S}
}

@BOOK{shibata16,
       author = {{Shibata}, Masaru},
        title = "{Numerical Relativity}",
         year = 2016,
          doi = {10.1142/9692},
    publisher = {World Scientific},
       adsurl = {https://ui.adsabs.harvard.edu/abs/2016nure.book.....S}
}

@ARTICLE{tanaka13a,
   author = {{Tanaka}, M. and {Hotokezaka}, K.},
    title = "{Radiative Transfer Simulations of Neutron Star Merger Ejecta}",
  journal = {ApJ},
archivePrefix = "arXiv",
   eprint = {1306.3742},
 primaryClass = "astro-ph.HE",
     year = 2013,
    month = oct,
   volume = 775,
      eid = {113},
    pages = {113},
      doi = {10.1088/0004-637X/775/2/113},
   adsurl = {http://adsabs.harvard.edu/abs/2013ApJ...775..113T}
}

@ARTICLE{tanaka20a,
       author = {{Tanaka}, Masaomi and {Kato}, Daiji and {Gaigalas}, Gediminas and {Kawaguchi}, Kyohei},
        title = "{Systematic opacity calculations for kilonovae}",
      journal = {MNRAS},
         year = 2020,
        month = aug,
       volume = {496},
       number = {2},
        pages = {1369-1392},
          doi = {10.1093/mnras/staa1576},
archivePrefix = {arXiv},
       eprint = {1906.08914},
 primaryClass = {astro-ph.HE},
       adsurl = {https://ui.adsabs.harvard.edu/abs/2020MNRAS.496.1369T}
}

@ARTICLE{thielemann26,
       author = {{Thielemann}, Friedrich-Karl and {Cowan}, John J.},
        title = "{The r-Process: History, Required Conditions, Astrophysical Sites, and Observations}",
      journal = {arXiv e-prints},
         year = 2026,
        month = jan,
          eid = {arXiv:2601.17246},
        pages = {arXiv:2601.17246},
          doi = {10.48550/arXiv.2601.17246},
archivePrefix = {arXiv},
       eprint = {2601.17246},
 primaryClass = {astro-ph.SR},
       adsurl = {https://ui.adsabs.harvard.edu/abs/2026arXiv260117246T}
}

@PHDTHESIS{winteler12,
   author = {{Winteler}, C.},
    title = "{Light Element Production in the Big Bang and the Synthesis of Heavy Elements in 3D MHD Jets from Core-Collapse Supernovae}",
   school = {University Basel, CH},
     year = 2012
}

@ARTICLE{wollaeger18a,
       author = {{Wollaeger}, Ryan T. and {Korobkin}, Oleg and {Fontes}, Christopher J. and
         {Rosswog}, Stephan K. and {Even}, Wesley P. and
         {Fryer}, Christopher L. and {Sollerman}, Jesper and
         {Hungerford}, Aimee L. and {van Rossum}, Daniel R. and
         {Wollaber}, Allan B.},
        title = "{Impact of ejecta morphology and composition on the electromagnetic signatures of neutron star mergers}",
      journal = {MNRAS},
         year = 2018,
        month = aug,
       volume = {478},
       number = {3},
        pages = {3298-3334},
          doi = {10.1093/mnras/sty1018},
archivePrefix = {arXiv},
       eprint = {1705.07084},
 primaryClass = {astro-ph.HE},
       adsurl = {https://ui.adsabs.harvard.edu/abs/2018MNRAS.478.3298W}
}

@ARTICLE{wu19,
       author = {{Wu}, Meng-Ru and {Barnes}, J. and {Martinez-Pinedo}, G. and {Metzger}, B.~D.},
        title = "{Fingerprints of Heavy-Element Nucleosynthesis in the Late-Time Lightcurves of Kilonovae}",
      journal = {Phys. Rev. Lett.},
         year = 2019,
        month = feb,
       volume = {122},
       number = {6},
          eid = {062701},
        pages = {062701},
          doi = {10.1103/PhysRevLett.122.062701},
archivePrefix = {arXiv},
       eprint = {1808.10459},
 primaryClass = {astro-ph.HE},
       adsurl = {https://ui.adsabs.harvard.edu/abs/2019PhRvL.122f2701W}
}

@ARTICLE{li98,
   author = {{Li}, L.-X. and {Paczy{\'n}ski}, B.},
    title = "{Transient Events from Neutron Star Mergers}",
  journal = {ApJL},
   eprint = {arXiv:astro-ph/9807272},
     year = 1998,
    month = nov,
   volume = 507,
    pages = {L59-L62},
      doi = {10.1086/311680},
   adsurl = {http://adsabs.harvard.edu/abs/1998ApJ...507L..59L}
}

@ARTICLE{kawaguchi24,
      title={Three dimensional end-to-end simulation for kilonova emission from a black-hole neutron-star merger}, 
      author={Kyohei Kawaguchi and Nanae Domoto and Sho Fujibayashi and Hamid Hamidani and Kota Hayashi and Masaru Shibata and Masaomi Tanaka and Shinya Wanajo},
journal={MNRAS},
year={2024},
      eprint={2404.15027},
      archivePrefix={arXiv},
      primaryClass={astro-ph.HE},
      url={https://arxiv.org/abs/2404.15027},
      }

@ARTICLE{kiuchi15,
   author = {{Kiuchi}, K. and {Cerd{\'a}-Dur{\'a}n}, P. and {Kyutoku}, K. and 
	{Sekiguchi}, Y. and {Shibata}, M.},
    title = "{Efficient magnetic-field amplification due to the Kelvin-Helmholtz instability in binary neutron star mergers}",
  journal = {Phys. Rev. D},
archivePrefix = "arXiv",
   eprint = {1509.09205},
 primaryClass = "astro-ph.HE",
     year = 2015,
    month = dec,
   volume = 92,
   number = 12,
      eid = {124034},
    pages = {124034},
      doi = {10.1103/PhysRevD.92.124034},
   adsurl = {http://cdsads.u-strasbg.fr/abs/2015PhRvD..92l4034K}
}

@ARTICLE{kiuchi17,
       author = {{Kiuchi}, Kenta and {Kawaguchi}, Kyohei and {Kyutoku}, Koutarou and {Sekiguchi}, Yuichiro and {Shibata}, Masaru and {Taniguchi}, Keisuke},
        title = "{Sub-radian-accuracy gravitational waveforms of coalescing binary neutron stars in numerical relativity}",
      journal = {Phys. Rev. D},
         year = 2017,
        month = oct,
       volume = {96},
       number = {8},
          eid = {084060},
        pages = {084060},
          doi = {10.1103/PhysRevD.96.084060},
archivePrefix = {arXiv},
       eprint = {1708.08926},
 primaryClass = {astro-ph.HE},
       adsurl = {https://ui.adsabs.harvard.edu/abs/2017PhRvD..96h4060K}
}

@ARTICLE{kiuchi18,
   author = {{Kiuchi}, K. and {Kyutoku}, K. and {Sekiguchi}, Y. and {Shibata}, M.
	},
    title = "{Global simulations of strongly magnetized remnant massive neutron stars formed in binary neutron star mergers}",
  journal = {Phys. Rev. D},
archivePrefix = "arXiv",
   eprint = {1710.01311},
 primaryClass = "astro-ph.HE",
     year = 2018,
    month = jun,
   volume = 97,
   number = 12,
      eid = {124039},
    pages = {124039},
      doi = {10.1103/PhysRevD.97.124039},
   adsurl = {http://adsabs.harvard.edu/abs/2018PhRvD..97l4039K}
}

@ARTICLE{kiuchi24,
       author = {{Kiuchi}, Kenta and {Reboul-Salze}, Alexis and {Shibata}, Masaru and {Sekiguchi}, Yuichiro},
        title = "{A large-scale magnetic field produced by a solar-like dynamo in binary neutron star mergers}",
      journal = {Nature Astronomy},
         year = 2024,
        month = mar,
       volume = {8},
        pages = {298-307},
          doi = {10.1038/s41550-024-02194-y},
archivePrefix = {arXiv},
       eprint = {2306.15721},
 primaryClass = {astro-ph.HE},
       adsurl = {https://ui.adsabs.harvard.edu/abs/2024NatAs...8..298K}
}

@ARTICLE{kulkarni05,
   author = {{Kulkarni}, S.~R.},
    title = "{Modeling Supernova-like Explosions Associated with Gamma-ray Bursts with Short Durations}",
  journal = {ArXiv Astrophysics e-prints},
   eprint = {arXiv:astro-ph/0510256},
     year = 2005,
    month = oct,
   adsurl = {http://adsabs.harvard.edu/abs/2005astro.ph.10256K}
}

@ARTICLE{kyutoku14,
       author = {{Kyutoku}, Koutarou and {Ioka}, Kunihito and {Shibata}, Masaru},
        title = "{Ultrarelativistic electromagnetic counterpart to binary neutron star mergers}",
      journal = {MNRAS},
         year = 2014,
        month = jan,
       volume = {437},
       number = {1},
        pages = {L6-L10},
          doi = {10.1093/mnrasl/slt128},
archivePrefix = {arXiv},
       eprint = {1209.5747},
 primaryClass = {astro-ph.HE},
       adsurl = {https://ui.adsabs.harvard.edu/abs/2014MNRAS.437L...6K}
}

@ARTICLE{combi23,
       author = {{Combi}, Luciano and {Siegel}, Daniel M.},
        title = "{GRMHD Simulations of Neutron-star Mergers with Weak Interactions: r-process Nucleosynthesis and Electromagnetic Signatures of Dynamical Ejecta}",
      journal = {ApJ},
         year = 2023,
        month = feb,
       volume = {944},
       number = {1},
          eid = {28},
        pages = {28},
          doi = {10.3847/1538-4357/acac29},
archivePrefix = {arXiv},
       eprint = {2206.03618},
 primaryClass = {astro-ph.HE},
       adsurl = {https://ui.adsabs.harvard.edu/abs/2023ApJ...944...28C}
}



\appendix
\renewcommand*{\sectionautorefname}{Appendix}

\section{Refining the NSE Temperature Limits}
\label{sec:appendix_B}

The transition temperature used before, i.e.\ the temperature at which we assume that the material leaves nuclear statistical equilibrium and where we start solving the full reaction network, was intentionally chosen conservatively large ($T_{\rm switch}=9\,\mathrm{GK}$). We find that the exact choice of $T_{\rm switch}$ can noticeably affect the nucleosynthesis outcome.

We show this in Fig.~\ref{fig:T_switch_finab}, where we plot the abundance patterns $Y(A)$ after $1$ year for two representative points in our parameter grid (both starting at $T_0=10\,\mathrm{GK}$ and expand with $\vel=0.3\,c$) for different values of $T_{\rm switch}$. For some points such as $(s_0,Y_{e,0})=(10.8\,\rm {k_B\,baryon}^{-1},0.153)$, the abundances remain very similar even when switching as low as $T_{\rm switch}=5\,\mathrm{GK}$ (and likely below). In contrast, we obtain strong deviations for other points in the parameter space, e.g.\ $(187\,\rm {k_B\,baryon}^{-1},0.418)$.


\begin{figure*}
  \centering
  \includegraphics[width=0.99\textwidth]{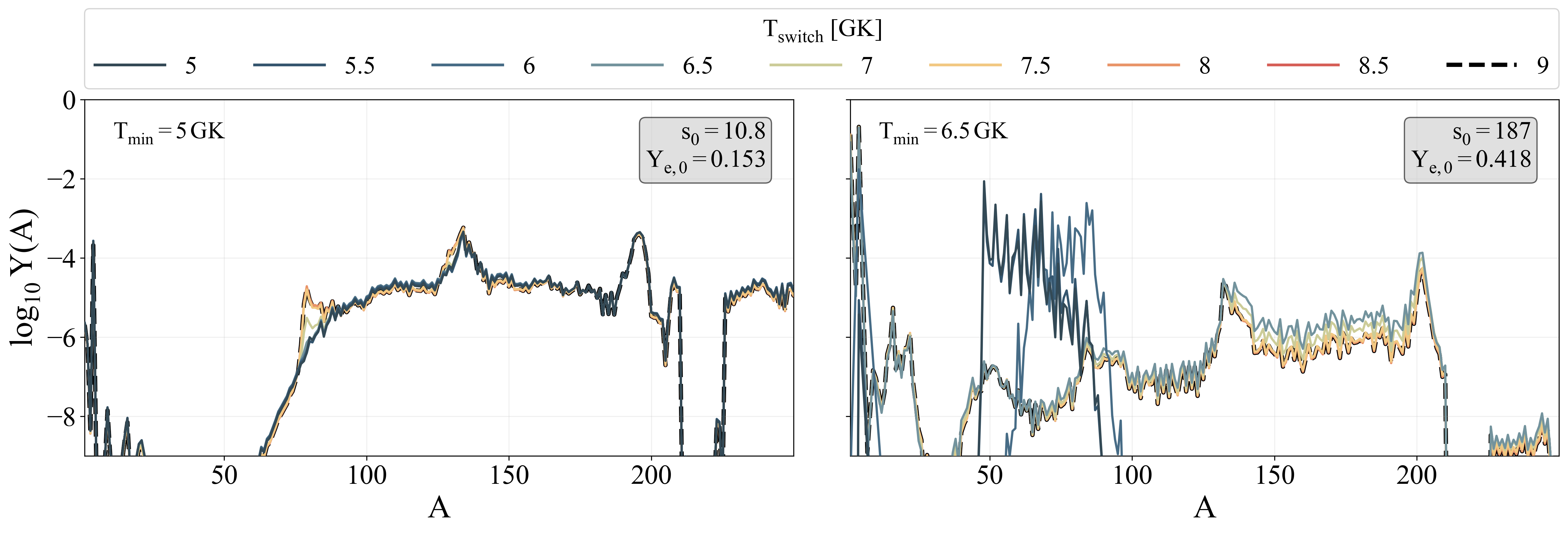}
  \caption{Final abundances after $1\,\mathrm{yr}$ for two representative trajectories from the $T_0=10\,\mathrm{GK}$ grid, shown for different NSE to network transition temperatures $T_{\rm switch}$. Colors indicate $T_{\rm switch}$; the dashed curve marks the reference choice $T_{\rm switch}=9\,\mathrm{GK}$. Left: $(s_0,Y_{e,0})=(10.8\,\rm {k_B\,baryon}^{-1},0.153)$, where the pattern is nearly unchanged down to $T_{\rm switch}=5\,\mathrm{GK}$. Right: $(s_0,Y_{e,0})=(187\,\rm {k_B\,baryon}^{-1},0.418)$, where lowering $T_{\rm switch}$ leads to pronounced differences.}
  \label{fig:T_switch_finab}
\end{figure*}


The origin of this sensitivity lies in the density-dependent seed-formation phase, where free nucleons cluster into light and intermediate-mass nuclei. During this stage, the neutron-to-seed ratio can change by orders of magnitude. In low-entropy outflows, seed injection remains efficient (see Sec.~\ref{sec:analytic_criterion}), keeping NSE valid down to low temperatures. In high-entropy ejecta (corresponding to low initial densities), these rates stall while NSE continues to track the shifting equilibrium to lower temperatures where it increasingly favors bound nuclei, driving $n_s$ below its true value and affecting the subsequent r-process.

To determine where NSE provides a good approximation over the relevant
temperature range, we computed an additional coarse grid of 120\,000 trajectories covering the same parameter space
as in Sec.~\ref{sec:trajectories}, but with varying $T_{\rm switch}$.
For each point $(s_0, Y_{e,0}, T_{\rm switch})$, we compare the final mass
fractions $X_{T_{\rm switch}}(A)$ to a reference
run $X_{\rm ref}(A)$ at $(s_0, Y_{e,0}, 9\,\mathrm{GK})$, assuming that the
system is safely in NSE above this temperature.
As a deviation measure we employ the total variation distance
\begin{equation}
    D \;=\; \frac{1}{2} \sum_A \left| X_{\rm ref}(A) - X_{T_{\rm switch}}(A) \right|\,,
    \label{eq:D}
\end{equation}
where $D \in [0,1]$ represents the fraction of the total baryonic mass
located at wrong mass numbers $A$ in the $T_{\rm switch}$ run relative
to the reference. We define $T_{\min}$ as the lowest $T_{\rm switch}$ for which $D < 0.5$, i.e.\ more than half of the total mass is correctly distributed. Figure~\ref{fig:T_switch_ps} shows $T_{\min}$ in the $s_0-Y_{e,0}$ plane for a velocity of $0.3\,c$. It represents the temperature above which NSE is a valid approximation and below which it breaks down.


\begin{figure}
  \centering
  \includegraphics[width=0.99\columnwidth]{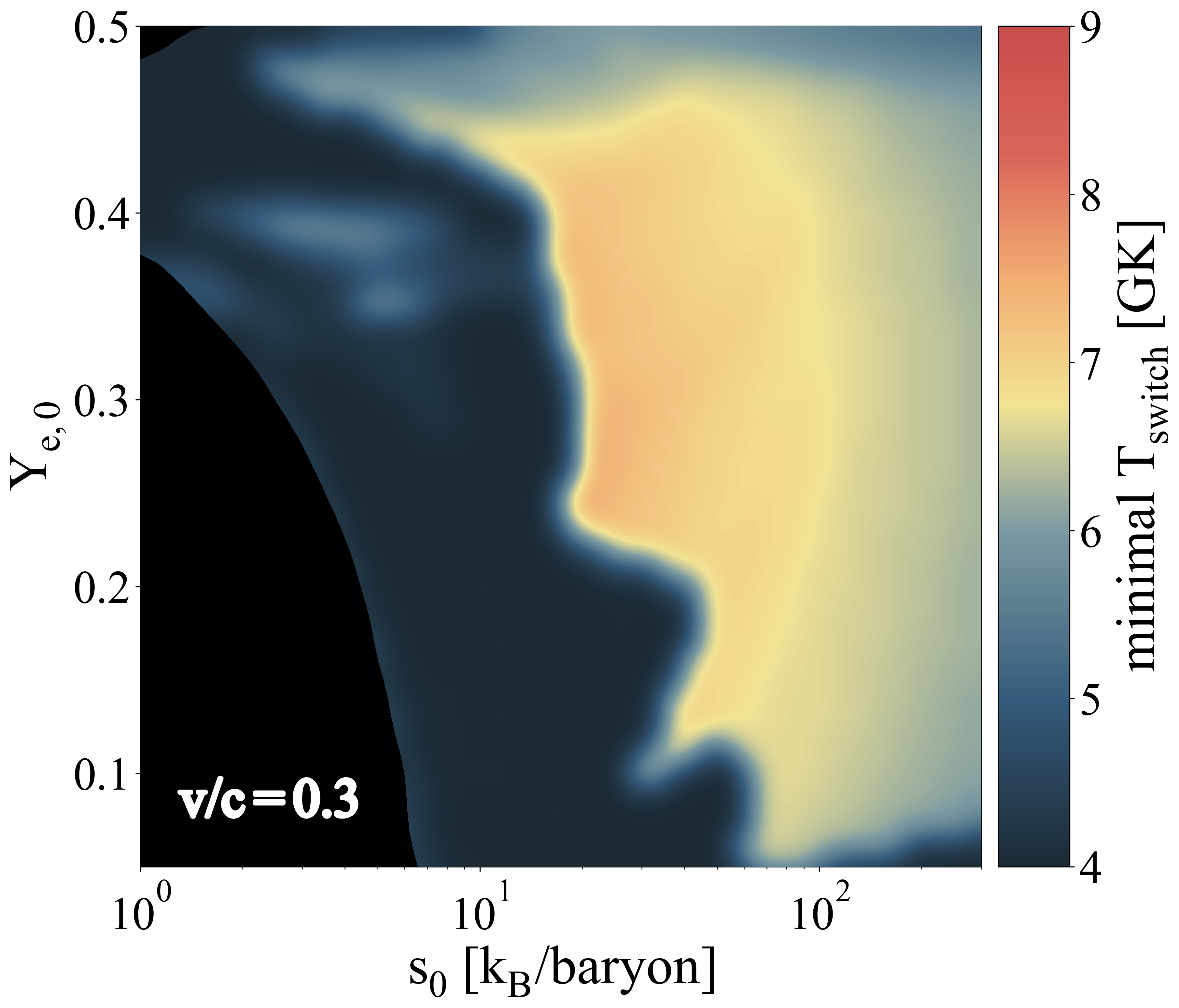}
    \caption{Minimum temperature $T_{\min}$ for which keeping the composition in NSE yields abundances consistent with a $T_{\rm switch}=9\,\mathrm{GK}$ reference run, evaluated on a coarse $(s_0,Y_{e,0})$ grid with $\vel=0.3\,c$. For each point we vary $T_{\rm switch}$ and compute the deviation measure $D$ (Eq.~(\ref{eq:D})); we then define $T_{\min}$ as the smallest $T_{\rm switch}$ with $D<0.5$.}
  \label{fig:T_switch_ps}
\end{figure}


\section{Sensitivity to the initial temperature}
\label{sec:appendix_A}

To explore how sensitive the nucleosynthesis
is to the starting temperature, we used the same entropies and electron fractions, but additionally explored initial temperatures of $T_0=0.5$, $2$, $4$, $6$, and $8\,\rm GK$. These simulations were performed on a coarser grid comprising $10\,000$ trajectories, with a fixed expansion velocity of $0.3\,c$. This allows us in particular to study tidally ejected material, which typically has lower initial temperatures and entropies and retains an electron fraction close to its original $\beta$-equilibrium value inside the isolated neutron star.

\citet{Freiburghaus1999} demonstrated that cold, neutron-rich matter can heat up to (or close to) NSE temperatures on extremely short timescales, provided the density is sufficiently high. This suggests robustness with respect to the initial temperature $T_0$, since once NSE is reached, the composition becomes largely insensitive to the details of the earlier evolution. 
Unfortunately, in Fig.~3 of the original paper \citep{Freiburghaus1999} the lines are hardly visible in the journal version of the paper, therefore we reproduce here the corresponding plot with the current version of \texttt{WinNet}, see
Fig.~\ref{fig:heating_into_NSE}. It  assumes a pure nucleon gas as the initial composition, using the full network at all temperatures. We find that the exact strength of this reheating depends on the initial density. Lowering $T_0$ at fixed $(s_0,Y_{e,0})$ reduces $\rho_0$, see Eq.~(\ref{eq:srad_closure_analytic}).
As a result, the expanding material reaches the regime where neutron captures become inefficient earlier and a larger fraction of neutrons remains unbound. Figure~\ref{fig:var_temp} shows the regions in parameter space where $X_n(1\,\rm min)>0.05$. A strong dependence on the initial electron fraction appears only at sufficiently high $T_0$, whereas at low $T_0$ the outcome is largely controlled by the entropy.


\begin{figure}
  \centering
  \includegraphics[width=0.99\columnwidth]{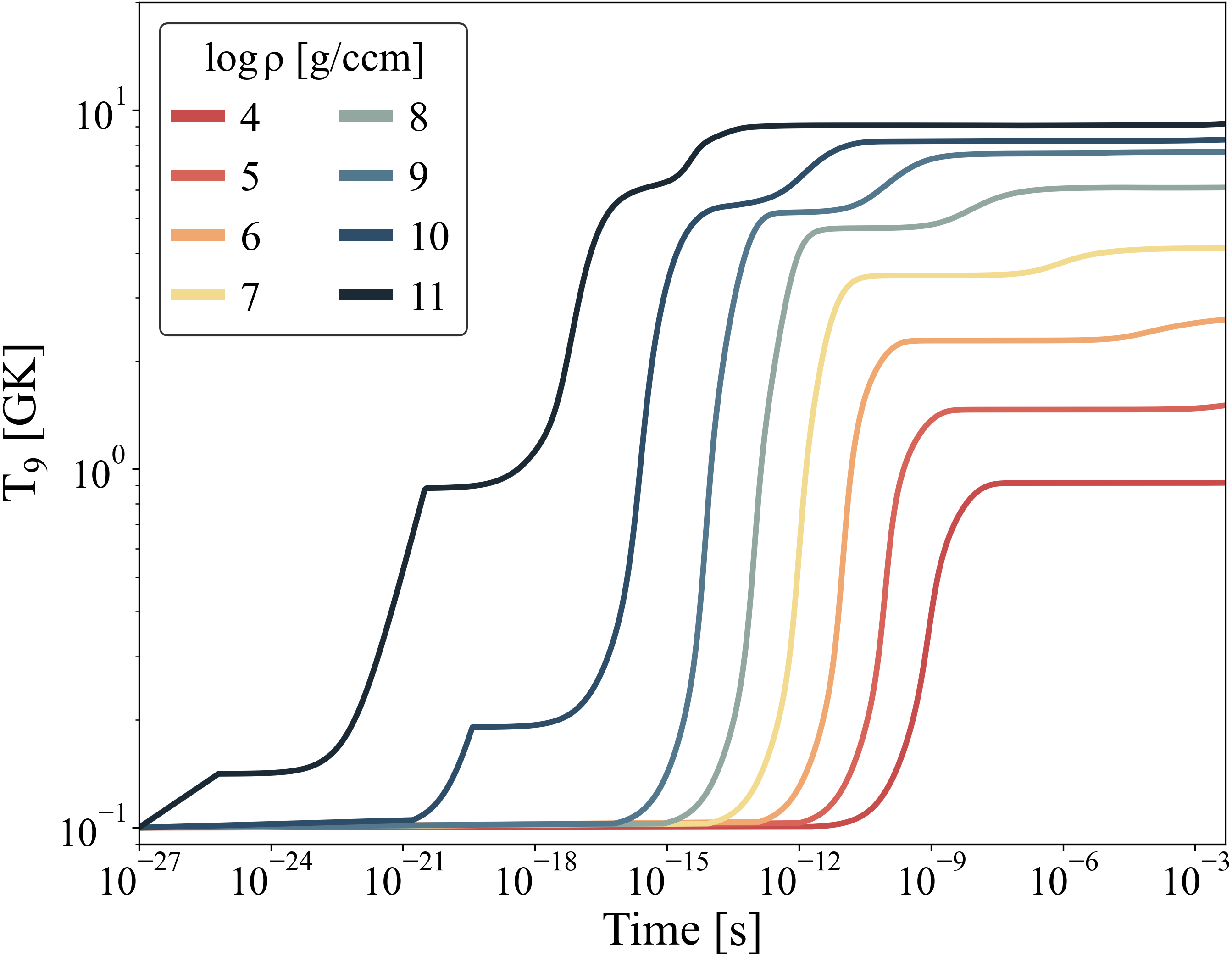}
  \caption{\textbf{Reproduction of Fig.~3 from} \citet{Freiburghaus1999}. Starting from a temperature of $T_0=0.1\,\rm GK$ for a pure nucleon gas with $Y_e=0.05$, the recombination of neutrons and protons into $\alpha$-particles leads to rapid heating. The effect is shown for different densities. Thus, even if the initial (high) temperatures in the hydrodynamic calculation are inaccurate, initially cold material will quickly heat up to temperatures where NSE is reached. This calculation is performed using the reaction network described in Section~\ref{sec:network}, \emph{without} activating the NSE-mode at temperatures above $9\,\rm GK$. Kinks in the curves can be traced back to individual bottleneck reactions, such as the formation of $^{2}\rm H$.}
  \label{fig:heating_into_NSE}
\end{figure}

%

\begin{figure}
  \centering
  \includegraphics[width=0.99\columnwidth]{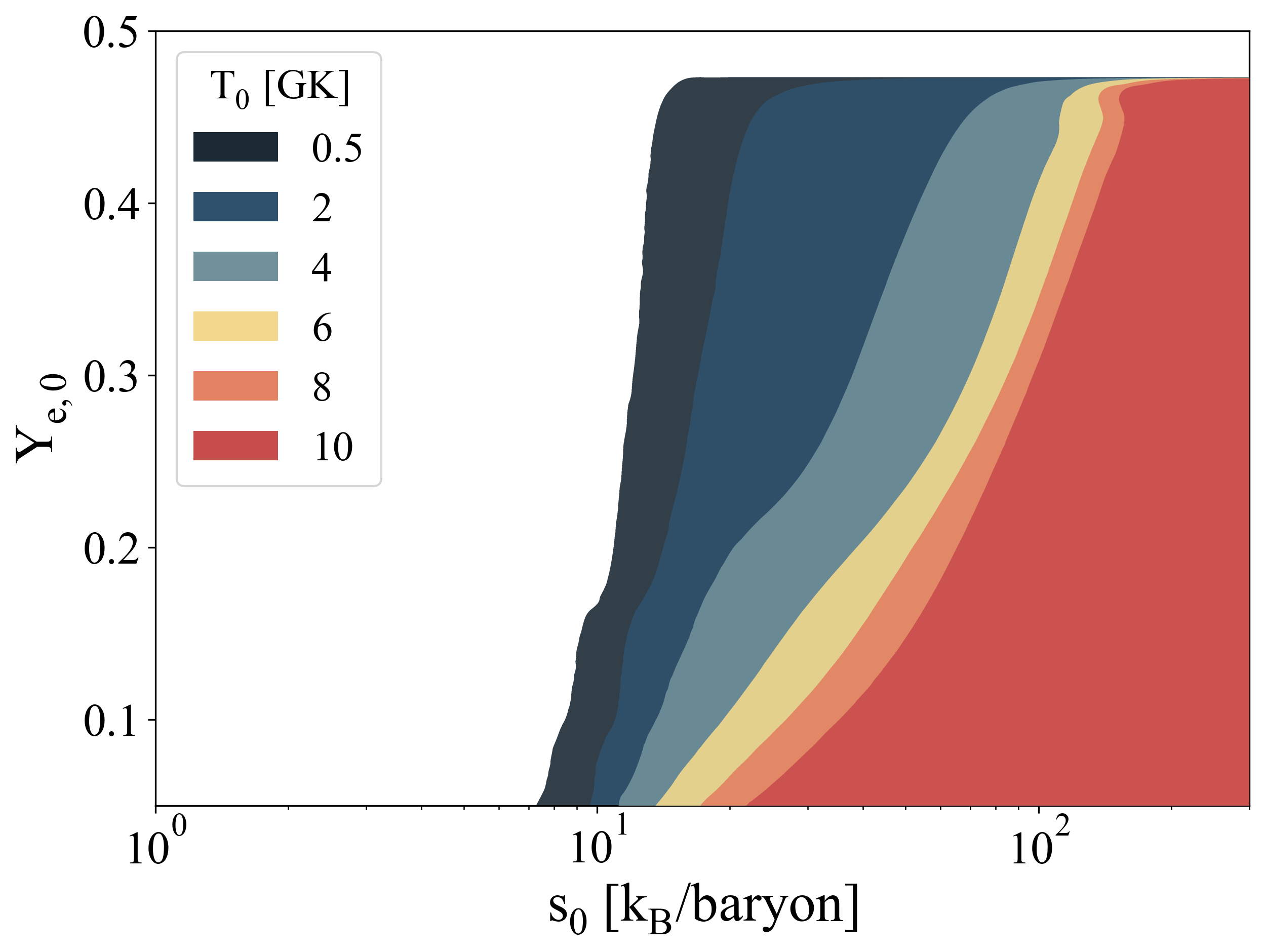}
  \caption{Areas in the parameter space with more than $5\,\%$ of the ejected mass in free neutrons (at $1\, \rm min$ after the merger) for various initial temperatures $T_0$ and a global outflow velocity of $0.3\, \rm c$, assuming an initial composition of free nucleons. }
  \label{fig:var_temp}
\end{figure}


\section{Thermalization efficiency of free neutron decay electrons}
\label{sec:appendix_C}
The electron produced in a free neutron decay can at most carry a kinetic energy of \(T = Q \approx 0.782\,\mathrm{MeV}\), where we measure temperatures in energy units, i.e. $k_B=1$. In reality, however, the decay energy is shared between the electron and the antineutrino. This is a consequence of the large mass difference between the proton and the electron and momentum conservation. On average, the electron receives an energy of \(\langle T \rangle = 0.59\, m_e c^2\), which corresponds to approximately \(39\%\) of the \(Q\)-value \citep{kulkarni05}. Due to the decay, the electron acquires a Lorentz factor of
$\gamma_e = 1+\langle T \rangle / (m_e c^2) \approx 1.59$ (i.e. a velocity of $\vel_e / c =(1 - 1/\gamma_e^2)^{1/2} \approx 0.778$).\\

To estimate the thermalization efficiencies (i.e. the fraction of the kinetic energy of the $e^-$ deposited in the surrounding medium), we have to evaluate the processes through which the electron loses energy to the environment. We follow \citet{Shenhar24} and include ionization losses \citep{Longair11}, plasma losses \citep{Solodov08}, and bremsstrahlung \citep{Longair11}.

We define the composition- and time-dependent total stopping power experienced by the decay electrons as their energy loss per unit column density $dX = \rho \,dx$
\begin{equation}
    \biggr(\frac{dE_e}{dX}\biggl)_{\rm tot} = \biggr(\frac{dE_e}{dX}\biggl)_{\bar A,\bar{Z}}^{\rm ion} +\, \biggr(\frac{dE_e}{dX}\biggl)_{\bar A,\bar{Z}}^{\rm plasma} +\, \biggr(\frac{dE_e}{dX}\biggl)_{\bar A,\bar{Z}}^{\rm brem}\,.
    \label{eq:total_stopping_power}
\end{equation}
With the effective average ionization potential \citep{Segre77} $\bar{I}=9.1\,\bar{Z}\bigl(1+1.9/\bar{Z}^{2/3}\bigr)\,\rm{eV}$ we can express the ionization contribution by
\begin{align}
    \biggr(\frac{dE_e}{dX}\biggl)_{\bar A,\bar{Z}}^{\rm ion}=\frac{4\pi e^4}{m_em_p \vel_e^2}\frac{\bar{Z}}{\bar{A}}\biggl[\ln{\biggl(\frac{\gamma_e^2m_e \vel_e^2}{\bar{I}}\biggr)}-\frac{\ln{\bigl(1+\gamma_e\bigr)}}{2}\\-\biggl(\frac{2\gamma_e+\gamma_e^2-1}{2\gamma_e^2}\biggr)\,\ln{2}+\frac{1}{2\gamma_e^2}+\frac{1}{16}\biggl(1-\frac{1}{\gamma_e}\biggr)^2\biggr]\,.
    \label{eq:ionization_stopping_power}
\end{align}
Plasma losses are accounted for by
\begin{align}
    \biggr(\frac{dE_e}{dX}\biggl)_{\bar A,\bar{Z}}^{\rm plasma}=\frac{2\pi e^4}{m_em_p \vel_e^2}\frac{\chi_e}{\bar{A}}\Biggl[\ln\Biggl({\biggl(\frac{E_e}{\hbar \omega_p}\biggr)^2\,\frac{\gamma_e+1}{2\gamma_e^2}}\Biggr)+1\\+\frac{1}{8}\biggl(\frac{\gamma_e-1}{\gamma_e}\biggr)^2-\biggl(\frac{2\gamma_e-1}{\gamma_e^2}\biggr)\,\ln{2}\Biggr]\,.
    \label{eq:plasma_stopping_power}
\end{align}
Here, we chose for simplicity that the number of free electrons per atom is $\chi_e\approx1$, so that the plasma frequency can be estimated by $\omega_p=\sqrt{4\pi e^2 n_e/m_e}$, with the free electron number density $n_e=\chi_e\,\rho/(\bar{A}\,m_p)$. In the vicinity of ions, high-energy electrons lose energy via bremsstrahlung, for which we use
\begin{align}
    \biggr(\frac{dE_e}{dX}\biggl)_{\bar A,\bar{Z}}^{\rm brems}=\frac{4e^4}{m_em_p \vel_ec}\frac{\bar{Z}^2e^2}{\bar{A}\hbar c}\frac{E_e}{m_ec^2}\biggl[\ln{\biggl(\frac{183}{\sqrt{\bar{Z}}}} \biggr)+\frac{1}{8}\biggr]\,.
    \label{eq:brems_stopping_power}
\end{align}
With the deposition energy-loss timescale \citep{Shenhar24}
\begin{equation}
    t_l\equiv \frac{E_e}{|\dot{E}_e|}=\frac{E_e}{\rho \, \vel_e \big(\frac{dE_e}{dX}\big)_{\rm tot}}\,,
\end{equation}
the thermalization efficiency of free neutron decay electrons can be estimated as 
\begin{equation}
    \epsilon_{\rm{th,n}}(t,E_e)\approx \begin{cases}
1 & ,\,\rm{for}\,\,t_l\leq t  \\
\frac{t}{t_l} & ,\,\rm{for}\,\,t_l\geq t\,.
\end{cases}
\end{equation}
We show the time, where full thermalization starts to break down ($\epsilon_{\rm{th,n}}<0.9$) in units of the mean lifetime of free neutrons $\tau_n$ in Fig.~\ref{fig:e_th,n} for our fastest ejecta models with $\vel=0.9\,c$ and electrons of the energy $E_e=\langle T\rangle$. We find that at the time when the neutrons decay, the electrons thermalize completely. Even in the most unfavorable case, we obtain $t(\epsilon_{\rm{th,n}}<0.9) / \tau_n \approx 6.3$. This result is robust against the assumption $\chi_e \approx 1$, as we find that ionization losses alone ensure full thermalization even a in neutral ejecta.


\begin{figure}
  \centering
  \includegraphics[width=0.99\columnwidth]{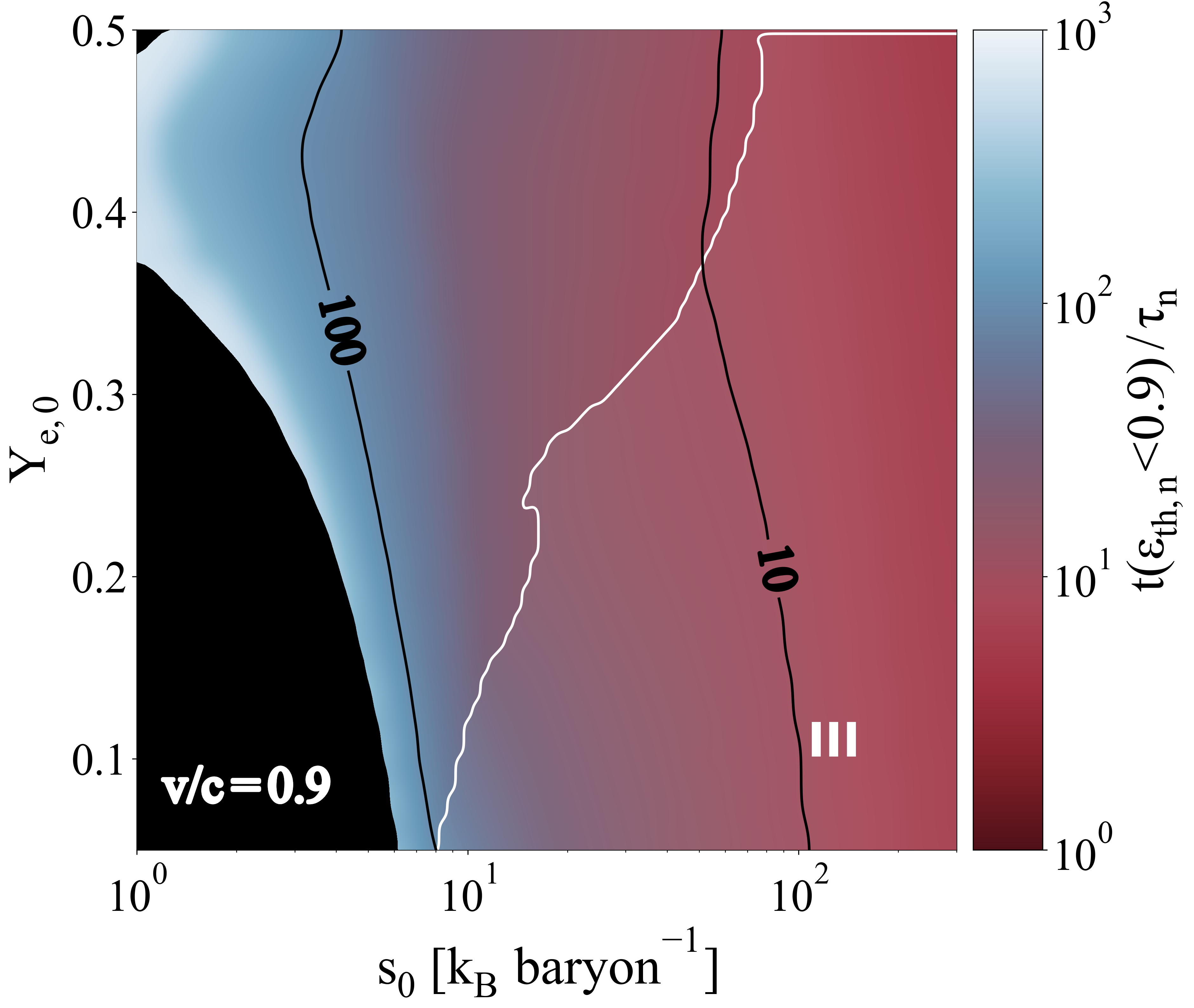}
    \caption{Time at which $\epsilon_{\rm th,n}<0.9$, normalized to the neutron lifetime $\tau_n$, for the fastest ejecta ($\vel=0.9\,c$) and electrons with $E_e=\langle T\rangle\approx 0.301\,\rm MeV$. The the white shaded region indicates the channel III trajectories where free neutrons actually occur. The black region marks the high-density parts of the parameter space, where the EOS is no longer valid.}
  \label{fig:e_th,n}
\end{figure}



\end{document}